\documentstyle[a4,12pt,epsf,axodraw]{article}
\begin{document}

\title{{\bf Mutual heavy ion dissociation in peripheral
collisions at ultrarelativistic energies}}
\author{
I.A. Pshenichnov$^{1,\star}$,  
J.P. Bondorf$^{2}$, I.N. Mishustin$^{2,3}$, \\ 
A. Ventura$^{4,\diamond}$,  S. Masetti$^{4,\dagger}$  \\
{\em $^1$ Institute for Nuclear Research, Russian Academy of Science,}\\
{\em 117312 Moscow, Russia}\\
{\em $^2$ Niels Bohr Institute, DK-2100 Copenhagen, Denmark}\\
{\em $^3$ Kurchatov Institute, Russian Research Center,}\\
{\em 123182 Moscow, Russia}\\
{\em $^4$ Italian National Agency for New Technologies,}\\
{\em Energy and the Environment, 40129 Bologna, Italy}\\
}

\date{ }
\maketitle

\begin{abstract}
We study mutual dissociation of heavy nuclei in 
peripheral collisions at ultrarelativistic energies. 
Earlier this process was proposed 
for beam luminosity monitoring via simultaneous registration 
of forward and backward neutrons in zero degree calorimeters at 
Relativistic Heavy Ion
Collider. Electromagnetic dissociation of heavy ions is considered 
in the framework of the Weizs\"{a}cker-Williams method and simulated by 
the RELDIS code. Photoneutron cross 
sections measured in different experiments and calculated by the GNASH code 
are used as input for the calculations of dissociation cross
sections.
The difference in results obtained with different inputs  
provides a realistic estimation for the systematic uncertainty of the 
luminosity monitoring method. Contribution to simultaneous neutron 
emission due to grazing nuclear interactions is calculated within
the abrasion model. Good description of CERN SPS experimental
data on Au and Pb dissociation gives confidence in predictive power of 
the model for AuAu and PbPb collisions at RHIC and LHC.
\end{abstract}

\noindent$^\star$ E-mail: pshenichnov@nbi.dk\\ 
\noindent$^\diamond$ E-mail: ventura@bologna.enea.it\\
\noindent$^\dagger$ ENEA guest researcher\\


\newpage

\section{Introduction}

The study of a new form of strongly interacting matter, 
the so-called quark-gluon plasma,
is at the core of current and future experimental programs at
Relativistic Heavy Ion Collider (RHIC) at Brookhaven 
National Laboratory (BNL)~\cite{RHIC} and Large Hadron Collider (LHC) at 
CERN~\cite{ALICE}. Although colliders give well-known advantages 
compared to the fixed target experiments, 
the kinematics of ultrarelativistic heavy-ion collisions at colliders 
creates certain complications in the beam monitoring 
as well as in the identification of collision events.      

Due to the geometrical factor $2\pi b$, where $b$ is the impact parameter,
the number of central nuclear collisions ($b\approx 0$) is relatively 
very small in the whole set of the collisions with nuclear overlap, 
$b\leq R_1+R_2$ ($R_1$ and $R_2$ are the nuclear radii).  
Moreover, in peripheral collisions without direct overlap of 
nuclear densities, $b>R_1+R_2$, one or both nuclei may be disintegrated 
by the long-range electromagnetic forces.
This process of Electromagnetic Dissociation (ED)
is a well-known phenomenon~\cite{Krauss,BaurHen}. 
The properties of central and peripheral collisions are very different 
and then should be studied separately. The ED 
events are less violent than the collisions with the 
participation of strong interactions. Namely, the average particle 
multiplicities are essentially lower~\cite{Pshenichnov2, Pshenichnov} and 
the main part of nucleons and mesons is produced in 
projectile and target fragmentation regions, very far from the 
mid-rapidity region.  
   
Calculations show~\cite{Krauss,Pshenichnov2,Weneser} that the ED cross section 
in collisions of heavy nuclei at RHIC and LHC by far exceeds the 
dissociation cross section due to the direct nuclear overlap.
In ${\rm AuAu}$ and ${\rm PbPb}$ collisions at such 
energies many neutrons can be produced in the ED 
process~\cite{Pshenichnov2}. Among other interesting phenomena one may  
expect a complete disintegration 
of nuclei induced by the electromagnetic fields of collision 
partners~\cite{Pshenichnov}. 
This phenomenon is very well known in nuclear reactions  
under the name of ``multifragmentation''~\cite{JPB}.   

Several operational problems of heavy-ion colliders are connected
with the high rate of the ED process. On the one hand, 
the ED process reduces the lifetimes of heavy ion beams in colliders as 
compared with the proton-proton accelerator mode~\cite{RHIC,ALICE,Weneser}.
On the other hand, the process of simultaneous neutron emission from the
collision partners, where the ED process plays a dominant role, 
can be useful for the luminosity 
monitoring~\cite{Baltz,White,Baltz2}.

The luminosity monitoring method based on mutual dissociation has several 
advantages~\cite{Baltz,White,Baltz2}. In particular, the
beam-residual-gas interaction events 
can be strongly suppressed in favour of the beam-beam events 
by the condition that 
a pair of neutrons should be detected in coincidence by each arm of the 
calorimeter. The cross section of mutual neutron emission can be 
calculated in the framework of conventional theoretical models 
designed for describing the heavy ion 
disintegration in 
peripheral collisions. Corresponding nuclear data, especially photoneutron
emission cross sections, may be used as numerical input
for such calculations. Therefore, the neutron counting rates in zero 
degree (very forward) calorimeters may provide an accurate measure 
for the heavy-ion collider luminosity. 

In the present paper the neutron emission in peripheral collisions of
ultrarelativistic heavy ions is considered with the aim of providing the 
theoretical basis for the luminosity monitoring 
method proposed in Refs.~\cite{Baltz,White,Baltz2}.  
The uncertainties in results originating from uncertainties 
in input nuclear data and in the theoretical model itself are 
carefully examined. 
A brief review of corresponding photonuclear data is given with 
special attention to the publications describing data evaluation and 
re-measurement. Model predictions for the ${\rm Au}$ and ${\rm Pb}$
fragmentation cross sections are compared with recent experimental 
data obtained in fixed target experiments at CERN SPS with the highest
energies available thus far. This serves as an important test before 
extrapolating our methods to the RHIC and LHC energies.

\section{Equivalent photon approach to simultaneous electromagnetic 
dissociation}\label{EMtheory}

\subsection{First order dissociation processes}

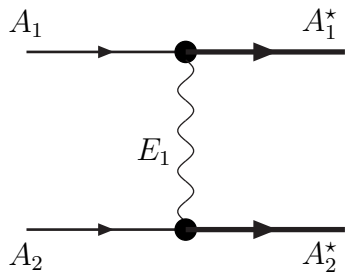
\begin{figure} 
\begin{center} 
\vspace{1cm}  
\unitlength=1cm 
\begin{picture}(4.0,2.0) 

\SetWidth{1.0}
\ArrowLine(0,100)(60,100)  
\Text(0,3.9)[]{$A_1$} 
\SetWidth{0.5}  
\GCirc(60,100){4}{0} 

\Photon(60,97)(60,36){3}{4} 
\Text(1.7,2.2)[]{$E_1$} 

\SetWidth{2.0}
\ArrowLine(60,100)(120,100)  
\Text(3.9,3.9)[]{$A_1^\star$}   

\SetWidth{1.0}
\ArrowLine(0,33)(60,33)  
\Text(0,0.8)[]{$A_2$} 
\SetWidth{0.5}  
\GCirc(60,33){4}{0} 

\SetWidth{2.0}
\ArrowLine(60,33)(120,33)  
\Text(3.9,0.8)[]{$A_2^\star$}   
  
\end{picture} 
\end{center} 
\caption{ 
Mutual electromagnetic excitation of relativistic nuclei: first order 
process. Closed circle denotes inelastic vertex.  
\label{EMfirst}  
} 
\end{figure}

The electromagnetic excitation of one of the collision partners, $A_2$, 
followed by its dissociation is schematically shown in Fig.~\ref{EM}. 
In such a process another partner, $A_1$, emits a photon, but remains 
in the ground state without any
nuclear excitation. Besides this ``classical'' process, one can consider 
a non-classical
process where the emission of a photon is accompanied by the nuclear 
excitation (see Fig.~\ref{EMfirst}), particularly, 
the giant resonance excitation. 
Such lowest-order contribution to the
simultaneous (mutual) excitation of the nuclei $A_1$ and $A_2$ was
considered in Refs.~\cite{Benesh,Hencken}. 
Also the correction to photon-photon luminosity function 
due to the inelastic photon emission was considered  
in Ref.~\cite{Hencken2} for $\gamma\gamma$ fusion reactions.

As shown in Refs.~\cite{Benesh,Hencken} (see also the discussion
in Ref.~\cite{BaurHen}), the lowest order process of simultaneous 
excitation of the collision partners has a small cross section.
For the cases of interest, i.e. for ${\rm AuAu}$ and ${\rm PbPb}$ collisions,
the cross sections for the simultaneous dipole-dipole excitation of such
nuclei are 0.49 and 0.54 mb, respectively~\cite{Benesh}. 
Using a rough estimation of Ref.~\cite{BaurHen}, $10^{-5}A^2$~mb,
one can get for the same nuclei 0.39 and 0.43 mb, respectively.
From the following discussion one will see that for heavy nuclei  
these first order contributions (Fig.~\ref{EMfirst}) are negligible 
compared to the second order ones (Fig.~\ref{EMsecond}). 
The latter leading order process with exchange of two photons 
is a classically allowed 
mechanism. It is considered in the next section where the 
formalism previously used in Ref.~\cite{Pshenichnov2} 
is extended to the case of mutual excitation.

\subsection{Second order dissociation processes}\label{EMD_MAIN}

Let us consider a collision of heavy ultrarelativistic nuclei at 
impact parameter $b>R_1+R_2$. The masses and charges of these nuclei 
are denoted as $A_1$, $Z_1$ and $A_2$, $Z_2$, respectively.
Hereafter the case of equal nuclei ($A_1=A_2=A$, $Z_1=Z_2=Z$ and
$R_1=R_2=R$) is investigated. Nevertheless, in some cases the indexes 
are used to show explicitly which of the collision partners emits or
absorbs photons.  
 
\begin{figure} 
\begin{center}
\vspace{1cm} 
\unitlength=1cm 
\begin{picture}(4.0,2.0) 

\SetWidth{1.0}
\ArrowLine(0,100)(60,100)  
\Text(0,3.9)[]{$A_1$} 
\SetWidth{0.5}  
\GCirc(60,100){4}{1} 

\Photon(60,97)(60,36){3}{4} 
\Text(1.7,2.2)[]{$E_1$} 

\SetWidth{1.0}
\ArrowLine(64,100)(120,100)  
\Text(3.9,3.9)[]{$A_1$}   

\SetWidth{1.0}
\ArrowLine(0,33)(60,33)  
\Text(0,0.8)[]{$A_2$} 
\SetWidth{0.5}  
\GCirc(60,33){4}{0} 

\SetWidth{2.0}
\ArrowLine(60,33)(120,33)  
\Text(3.9,0.8)[]{$A_2^\star$}   
  
\end{picture} 
\end{center} 
\caption{ 
Electromagnetic excitation of one of the colliding nuclei: first order 
process. Open and closed circles denote elastic and inelastic vertices,
respectively. 
\label{EM}  
} 
\end{figure}
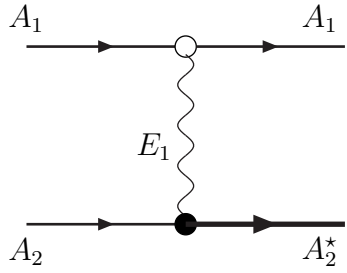

According to the Weizs\"{a}cker-Williams (WW) method~\cite{JDJackson}, 
the impact of the Lorentz-boosted Coulomb field of the nucleus $A_1$ 
on $A_2$ is treated as the absorption of an  
equivalent photon by the nucleus $A_2$ (see Fig.~\ref{EM}).     
In the rest frame of this nucleus the spectrum of virtual photons 
from the collision partner $A_1$ at impact parameter $b$ is expressed as: 
\begin{equation}
N_{Z_1}(E_1 ,b)=\frac{\alpha Z^{2}_1}{\pi ^2}
\frac{{\sf x}^2}{\beta ^2 E_1 b^2} 
\Bigl(K^{2}_{1}({\sf x})+\frac{1}{\gamma ^2}K^{2}_{0}({\sf x})\Bigl).
\label{WWspectrum}
\end{equation}
Here $\alpha$ is the fine structure constant, 
${\sf x}=E_1 b/(\gamma \beta \hbar c)$ is an argument of the modified Bessel 
functions of zero and first orders, $K_0$ and $K_1$, $\beta =v/c$ and 
$\gamma =(1-\beta^2)^{-1/2}$ is the Lorentz factor of the moving charge 
$Z_1$. If the Lorentz factor of each heavy-ion beam is $\gamma_{beam}$, then 
$\gamma= 2\gamma^2_{beam} -1$ for the case of collider. Hereafter the
natural units are used with $\hbar =c=1$.
   
The mean number of photons absorbed by the nucleus $A_2$ 
in the collision at impact parameter $b$ is defined by:
\begin{equation}
m_{A_2}(b)=
\int\limits_{E_{min}}^{E_{max}}N_{Z_1}(E_1 ,b)\sigma_{A_2}(E_1)dE_1 ,
\label{mb}
\end{equation}
where the appropriate total photoabsorption cross section, 
$\sigma_{A_2}(E_1)$ is used. For $E_{min}$ one usually takes
the neutron emission threshold, while
the upper limit of integration is $E_{max}\approx\gamma/R$.  
We assume that the probability of multiphoton absorption is given 
by the Poisson 
distribution with the mean multiplicity $m_{A_2}(b)$ defined by Eq.~(\ref{mb}).

Following Refs.~\cite{Pshenichnov2,Llope,BertBaur},
we express the cross section for the electromagnetic 
dissociation 
of {\em one} of the nuclei due to the absorption of a {\em single} photon
(Fig.~\ref{EM}) leading to 
a certain dissociation channel $i$ as:
\begin{equation}
\sigma^{ED}_1(i)=2\pi\int\limits_{b_{c}}^{\infty} bdb P_{A_2}(b),
\label{SS}
\end{equation}
where the probability of dissociation at impact parameter $b$ is defined as:  
\begin{equation}
P_{A_2}(b)=e^{-m_{A_2}(b)}\int\limits_{E_{min}}^{E_{max}} dE_1 N_{Z_1}(E_1,b)
\sigma_{A_2}(E_1)f_{A_2}(E_1,i),
\label{P1}
\end{equation}
and $f_{A_2}(E_1,i)$ is the branching ratio for the considered channel $i$ in
the absorption of a photon with energy $E_1$ on nucleus $A_2$. 
The choice of a critical impact parameter $b_c$, which separates 
the domains of nuclear and electromagnetic interactions, 
will be discussed in Sec.~\ref{parameters}.   

Let us turn now to the mutual dissociation process shown in 
Fig.~\ref{EMsecond}.
The corresponding graph may be constructed from two graphs of the single 
dissociation by interchanging the roles of ``emitter'' and 
``absorber'' at the secondary photon exchange. Several assumptions have 
to be made to obtain an expression for the mutual dissociation
cross section. 

Firstly, we suppose that the emitted photon 
with energy $E_1\leq E_{max}$ does not change essentially    
the total energy, $E_A=\gamma M_A$, of the emitting nucleus, where 
$M_A$ is the nuclear mass. This can be justified by estimating the ratio
\begin{equation}
 r=\frac{E_{max}}{E_A}\approx\frac{1}{R M_A}, 
\end{equation}
which is close to $10^{-4}$ for heavy nuclei. Therefore, the kinematical
conditions for the secondary photon exchange are very similar to those for
the primary one, and there are no correlations between the energies 
of the primary and secondary photons, $E_1$ and $E_2$. In other words, 
the primary and secondary photon exchanges may be considered 
as independent processes even if they take place in the same collision during
a short-term overlap of the Lorentz-contracted Coulomb fields of 
the colliding nuclei. Secondly, the equivalent photon spectrum from
the excited nucleus, $A_2^\star$ in the notations of Fig.~\ref{EMsecond}, is 
the same as the spectrum from the nucleus in its ground state, $A_2$.
This follows from the fact that at ultrarelativistic energies the 
collision time is much shorter then the characteristic deexcitation time
during
which a nucleus changes its initial charge via proton emission or fission.  

Following these assumptions, one can express the cross section
for the {\em mutual}
dissociation of nuclei $A_1$ and $A_2$ (Fig.~\ref{EMsecond}) to channels $i$ 
and $j$, respectively, as:
\begin{equation}
\sigma^{ED}_m(i\mid j)=2\pi\int\limits_{b_{c}}^{\infty} bdb 
P_{A_1}(b) P_{A_2}(b).
\label{SM}
\end{equation}
Substituting Eq.~(\ref{P1}) for each of the nuclei and changing the order of 
integration one obtains:
\begin{equation}
\sigma^{ED}_m(i\mid j)=
\int\limits_{E_{min}}^{E_{max}}\int\limits_{E_{min}}^{E_{max}}dE_1dE_2 
{\cal N}_m(E_1,E_2)\sigma_{A_2}(E_1)\sigma_{A_1}(E_2) 
f_{A_2}(E_1,i) f_{A_1}(E_2,j),
\label{SMFULL}
\end{equation}
where the spectral function ${\cal N}_m(E_1,E_2)$ for mutual dissociation 
is introduced:
\begin{equation}
{\cal N}_m(E_1,E_2)=2\pi\int\limits_{b_{c}}^{\infty} bdb 
e^{-2m(b)} N_{Z_1}(E_1,b) N_{Z_2}(E_2,b).  
\label{NM}
\end{equation}
Conditions $A_1=A_2=A$ and $Z_1=Z_2=Z$ were used in 
Eqs.~(\ref{SMFULL}) and (\ref{NM}), as it is usually in heavy-ion colliders,
and therefore $m_{A_1}(b)=m_{A_2}(b)=m(b)$. 
Nevertheless, the dissociation channels $i$ and $j$ may be different for 
each of the nuclei even in such case.

Several remarks may be made concerning Eqs.~(\ref{SMFULL}) and (\ref{NM}).
Compared to another process of the second order dissociation of a single 
nucleus (see Fig.~\ref{EM2}) some points of similarity may be found. 
Indeed, expressions given in Ref.~\cite{Pshenichnov2} 
for the corresponding cross section $\sigma^{ED}_2(i)$ of the
second order process are as following:
\begin{equation}
\sigma^{ED}_2(i)=
\int\limits_{E_{min}}^{E_{max}}\int\limits_{E_{min}}^{E_{max}}dE_1dE_2 
{\cal N}_2(E_1,E_2)\sigma_{A_2}(E_1)\sigma_{A_2}(E_2)
f_{A_2}(E_1,E_2,i),
\label{SD}
\end{equation}
\begin{equation}
{\cal N}_2(E_1,E_2)=\pi\int\limits_{b_{c}}^{\infty} bdb 
e^{-m(b)} N_{Z_1}(E_1,b) N_{Z_1}(E_2,b).  
\label{ND}
\end{equation}
However, there is an important difference in the definitions of branching 
ratios $f_{A_2}(E_1,i)$, $f_{A_1}(E_2,j)$ compared with $f_{A_2}(E_1,E_2,i)$, 
since the former are for the absorption of two photons by {\em two different}
nuclei leading to certain dissociation channels $i$ and $j$, while the 
latter is for the absorption of two photons by
a {\em single} nucleus leading to a channel $i$. 
Another difference is due to an additional
factor of $2e^{-m(b)}$ in Eq.~(\ref{SMFULL}) compared with Eq.~(\ref{SD}).
It comes from the fact that Eq.~(\ref{SMFULL}) contains 
the product of the Poisson probabilities for
the single photon absorption for the
collision with impact parameter $b$:
\begin{equation} 
P_{A_1}(b)P_{A_2}(b)= m^2(b) e^{-2m(b)}= 2e^{-m(b)} P^{(2)}_A(b),  
\end{equation}
while $P^{(2)}_A(b) = m^2(b) e^{-m(b)}/2$ is the Poisson probability for 
the double photon absorption~\cite{Pshenichnov2,Llope}.
Results for partial cross sections of mutual dissociation will be 
presented and discussed in Sec.~\ref{LHC}.

\subsection{Next-to-leading-order processes of mutual 
dissociation}\label{EMNLO}

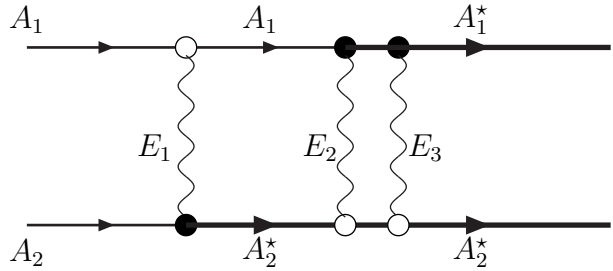
\begin{figure}
\begin{center} 
\vspace{1cm} 
\unitlength=1cm 
\begin{picture}(8.0,2.0) 

\SetWidth{1.0}
\ArrowLine(0,100)(60,100)  
\Text(0,3.9)[]{$A_1$} 
\SetWidth{0.5}  
\GCirc(60,100){4}{1.} 
\SetWidth{1.0}
\ArrowLine(64,100)(120,100)  
\Text(3.1,3.9)[]{$A_1$}   

\SetWidth{0.5}
\Photon(60,97)(60,36){3}{4} 
\Text(1.7,2.2)[]{$E_1$} 

\SetWidth{1.0}
\ArrowLine(0,33)(60,33)  
\Text(0,0.8)[]{$A_2$} 
\SetWidth{0.5}  
\GCirc(60,33){4}{0} 
\SetWidth{2.0}
\ArrowLine(60,33)(120,33)  
\Text(3.1,0.8)[]{$A_2^\star$}
\SetWidth{0.5}  
\GCirc(120,100){4}{0}
\SetWidth{2.0} 
\ArrowLine(120,33)(220,33)  
\Text(5.9,0.8)[]{$A_2^\star$}
\SetWidth{0.5}  
\GCirc(120,33){4}{1.} 
  
\Photon(120,97)(120,36){3}{4} 
\Text(3.9,2.2)[]{$E_2$}

\GCirc(140,100){4}{0}
\Photon(140,97)(140,36){3}{4} 
\Text(5.3,2.2)[]{$E_3$}
\GCirc(140,33){4}{1.}

\SetWidth{2.0}
\ArrowLine(120,100)(220,100)  
\Text(5.9,3.9)[]{$A_1^\star$}   
 
\end{picture} 
\end{center} 
\caption{ 
Mutual electromagnetic excitation of relativistic nuclei: 
next-to-leading-order contribution with single and double photon
exchange processes. Open and closed circles denote 
elastic and inelastic vertices,
respectively.  
\label{EMNLO12}  
} 
\end{figure}

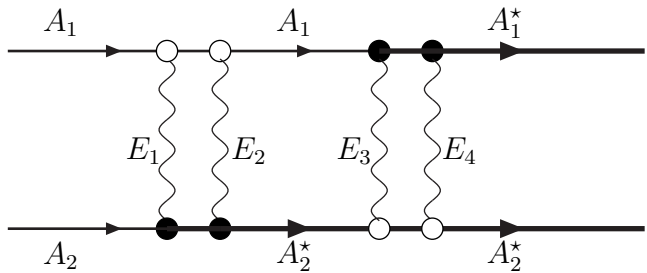
\begin{figure}
\begin{center} 
\vspace{1cm} 
\unitlength=1cm 
\begin{picture}(8.0,2.0) 

\SetWidth{1.0}
\ArrowLine(-20,100)(60,100)  
\Text(0,3.9)[]{$A_1$} 
\SetWidth{0.5}  
\GCirc(60,100){4}{1.} 
\SetWidth{1.0}
\ArrowLine(64,100)(120,100)  
\Text(3.1,3.9)[]{$A_1$}   

\SetWidth{0.5}
\Photon(60,97)(60,36){3}{4}
\Text(1.1,2.2)[]{$E_1$} 
\Photon(40,97)(40,36){3}{4} 
\GCirc(40,100){4}{1.}
\GCirc(40,33){4}{0}
\Text(2.5,2.2)[]{$E_2$} 

\SetWidth{1.0}
\ArrowLine(-20,33)(60,33)  
\Text(0,0.8)[]{$A_2$} 
\SetWidth{0.5}  
\GCirc(60,33){4}{0} 
\SetWidth{2.0}
\ArrowLine(60,33)(120,33)
\Line(40,33)(60,33)  
\Text(3.1,0.8)[]{$A_2^\star$}
\SetWidth{0.5}  
\GCirc(120,100){4}{0}
\SetWidth{2.0} 
\ArrowLine(120,33)(220,33)  
\Text(5.9,0.8)[]{$A_2^\star$}
\SetWidth{0.5}  
\GCirc(120,33){4}{1.} 
  
\Photon(120,97)(120,36){3}{4} 
\Text(3.9,2.2)[]{$E_3$}

\GCirc(140,100){4}{0}
\Photon(140,97)(140,36){3}{4} 
\Text(5.3,2.2)[]{$E_4$}
\GCirc(140,33){4}{1.}

\SetWidth{2.0}
\ArrowLine(120,100)(220,100)  
\Text(5.9,3.9)[]{$A_1^\star$}   
 
\end{picture} 
\end{center} 
\caption{ 
Mutual electromagnetic excitation of relativistic nuclei: 
next-to-leading-order contribution with two double photon
exchange processes. Open and closed circles denote 
elastic and inelastic vertices,
respectively.  
\label{EMNLO22}  
} 
\end{figure}

The second order process, which is shown in Fig.~\ref{EMsecond} and considered
in Sec.~\ref{EMD_MAIN}, is the leading order mechanism of mutual 
electromagnetic dissociation. This is confirmed by the calculations of the
corresponding total cross sections for the leading order (LO), 
Fig.~\ref{EMsecond} and next-to-leading-order (NLO) processes, 
Figs.~\ref{EMNLO12} and \ref{EMNLO22}.

Following the assumption that the probability of multiphoton absorption 
is given
by the Poisson distribution with the mean multiplicity $m_A(b)$ 
of Eq.(\ref{mb}), one has for the LO process shown in Fig.~\ref{EMsecond}:
\begin{equation}  
\sigma^{ED}_m({\rm LO})=2\pi\int\limits_{b_{c}}^{\infty} bdb 
m_A^2(b) e^{-2m_A(b)}. 
\label{LO}
\end{equation}
\noindent Here the case of equal masses and charges of 
collision partners is considered,
i.e. $A_1=A_2=A$ and $Z_1=Z_2=Z$.

For the NLO process with exchange of three photons (${\rm NLO_{12}}$), 
Fig.~\ref{EMNLO12}, the total
cross section is given by:
\begin{equation}  
\sigma^{ED}_m({\rm NLO_{12}})=2\pi\int\limits_{b_{c}}^{\infty} bdb 
\frac{m_A^3(b)}{2} e^{-2m_A(b)}. 
\label{NLO12}
\end{equation}
\noindent A complementary process (${\rm NLO_{21}}$) with the excitation 
of another nucleus, $A_2$, via double photon absorption is equally possible 
and has the same cross section.

Another NLO process of mutual dissociation is due to exchange of four
photons (${\rm NLO_{22}}$), Fig.~\ref{EMNLO22}, and its cross 
section can be written as:
\begin{equation}  
\sigma^{ED}_m({\rm NLO_{22}})=2\pi\int\limits_{b_{c}}^{\infty} bdb 
\frac{m_A^4(b)}{4} e^{-2m_A(b)}. 
\label{NLO22}
\end{equation}

Finally, one can calculate the sum of all contributions using 
the prescription of Ref.~\cite{Baltz2}:
\begin{equation}
\sigma^{ED}_m({\rm tot})=2\pi\int\limits_{b_{c}}^{\infty} bdb 
\biggl(1-e^{-m_A(b)}\biggr)^2. 
\label{SMTOT}
\end{equation}
\noindent Since for each of the ions the probability of collision 
without photon exchange is equal to $e^{-m_A(b)}$, 
Eq.(\ref{SMTOT}) is evident.

Calculations of $\sigma^{ED}_m({\rm tot})$, $\sigma^{ED}_m({\rm LO})$, 
$\sigma^{ED}_m({\rm NLO})$ and cross sections for 
specific dissociation channels, $\sigma^{ED}_m(i\mid j)$, with
and without NLO corrections were 
performed by the modified RELDIS code, which contains now a special 
simulation mode for the mutual electromagnetic dissociation process.
Our results for the total dissociation cross sections in 
electromagnetic and nuclear interactions are given in Tab.~\ref{T1}.
These cross sections values are in good agreement with results of 
other authors. 

The total cross sections for mutual electromagnetic dissociation 
given in Tab.~\ref{T1} are much lower than the cross sections 
for single dissociation.
However, even the former values are found to be comparable 
to the total nuclear dissociation cross sections, see Tab.~\ref{T1}.

As one can see, the ratios between the first and second order
processes are very different for single and mutual dissociation.
The first order dissociation process 
(with exchange of a single photon) can be safely neglected in
considering the mutual dissociation of heavy nuclei 
at ultrarelativistic energies.

Tab.~\ref{T1} contains also the cross sections for LO
process, $\sigma^{ED}_m({\rm LO})$. As one can see, the LO mechanism
gives $\sim 70$\% and $\sim 60$\% of the $\sigma^{ED}_m({\rm tot})$ 
at RHIC and LHC energies, respectively. The sum of the NLO contributions
to the total cross section gives additional
$\sim 25$--27\%. Therefore, at RHIC energies, for example, 
the remaining contribution of $\sim 5$\% (or $\sim 0.2$ b) 
is due to exotic triple excitations.

Finally, $\sigma_m^{ED}$ and ${\cal N}_m$ written for the LO
mutual dissociation process, Eqs.~(\ref{SMFULL}) and (\ref{NM}), can be
easily generalized to the case of different NLO processes. 
However, the resulting expressions are lengthy and we do not give it here
for brevity's sake.

\section{Abrasion model for mutual dissociation in
nuclear collisions}\label{abrmodel}

Several nucleons can be abraded from collision partners
in grazing nuclear collisions. We are interested in a situation when 
only few nucleons are removed. This is the case when
nuclear densities overlap weakly and mainly 
nuclear periphery is involved in the interaction.

The cross section for the abrasion of $a$ nucleons from the 
projectile $(A_1,Z_1)$ in a collision with 
the target $(A_2,Z_2)$ may be derived
from the Glauber multiple scattering theory~\cite{Hufner}: 
\begin{equation}
\sigma^{nuc}(a)={A_1\choose{a}}\times 2\pi
\int\limits_{0}^{b_{c}} bdb
\Bigl(1-P(b)\Bigr)^{a}P(b)^{A_1-a}.
\label{abr_main}
\end{equation}
Here $P(\vec{b})$ is calculated as the overlap of projectile, 
$\rho_{1}(\vec{r})$, and target $\rho_{2}(\vec{r})$
densities in the collision with impact parameter $b$:
\begin{equation}
P(\vec{b})=\int d^2\vec{s}D_1(\vec{s})
exp\Bigl(-A_2\sigma_{NN}D_2(\vec{s}+\vec{b})\Bigr),
\end{equation}
where the nuclear thickness functions,
\begin{equation}
D_{1,2}(\vec{s})=\int\limits_{-\infty}^{+\infty}d{\tt z}
\rho_{1,2}(\vec{s},{\tt z}),
\end{equation}
are introduced. 
In our calculations the nuclear density functions are
approximated by Fermi functions:
\begin{equation}
\rho_{1,2}(r)={\rho_o\over1+exp\Bigl({r-r_oA_{1,2}^{1/3}\over d}\Bigr)},
\label{Fermi}
\end{equation}
where $r_o$ is a parameter which defines the nuclear half-density 
radius, $R_o=r_o\times A^{1/3}$, and  
$d=0.54$ fm is the diffuseness parameter.  
The choice of these and other important 
parameters of the model, the integration cutoff parameter, $b_{c}$, 
and the total nucleon-nucleon cross section, $\sigma_{NN}$, is discussed 
in Sec.~\ref{parameters}.     

The above expressions determine only the number of nucleons 
removed from the projectile and do not specify how many
protons or neutrons were knocked out. 
Further assumptions are needed 
to determine the charge-to-mass ratio of the residual nucleus and hence
the numbers of protons, $z$, and neutrons, $n$, abraded from the 
initial nucleus (see also the discussion in Ref.~\cite{Oliveira,Gaimard}). 

In the present work we use the so-called hypergeometrical 
model~\cite{Oliveira,Gaimard}, assuming that each removed projectile 
nucleon has a $N_1/A_1$ probability ($N_1=A_1-Z_1$) of being a neutron:
\begin{equation}
\sigma^{nuc}(n,z)=\frac{ {Z_1\choose{z}} {N_1\choose{n}} } 
{{A_1\choose{a}}}\sigma^{nuc}(a).
\label{hyper}
\end{equation} 
In other words this means that there is no correlation at all between
the proton and neutron distributions and the abrasion process removes
protons and neutrons from the projectile nucleus in a random way.

Several physical processes, which might be important in 
heavy-ion collisions, were neglected in this model. 
The excited residual nucleus created due to the abrasion process should 
undergo its de-excitation on the second ablation step.
On this step more neutrons may be emitted via evaporation.
However, as it was noticed in Ref.~\cite{Oliveira}, the excitation
energies obtained on abrasion step due to removal of one or two nucleons
are generally not sufficient for intensive particle evaporation. 
Therefore, for the cases of interest, i.e. $1n$ and $2n$ dissociation
channels, the ablation step can be neglected.   
 
The abrasion of nucleons from projectile and target
proceeds via high-energy collisions between nucleons.   
Nucleon-antinucleon pairs may be created in such interactions
and neutrons may be presented in these pairs. However, as one
can find in a compilation~\cite{Gazdzicki},
even at high energies $\sim 100$ AGeV, the relative rate of such pair
production is not so high, $\sim 5\%$. Because of this, we do
not consider such processes in calculations of the neutron 
emission cross sections.  
 
Knocked-out nucleons can also suffer a final state interaction with
spectators~\cite{Oliveira}. We believe that this process is less 
important at high energies compared with 
intermediate energies of $\sim 0.1-1$ AGeV. 
For the latter case the escape probability is estimated to be 
$P_{esc}\sim 0.5-0.75$ for peripheral collisions of 
heavy nuclei~\cite{BenCook}. The momenta of recoil nucleons
may be comparable with the Fermi momentum of intranuclear nucleons and
their angular distribution is very wide so that they can 
be easily captured by one of the spectators.  
The situation is different at high energies, where the 
transverse momenta of collided nucleons are typically large,
of order of 0.5 GeV, and therefore their
subsequent capture is less probable.
Other effects like a finite hadronization length may further reduce
the secondary interaction probability.
Therefore, we assume that in peripheral collisions at RHIC and LHC 
the probability for each of the collided nucleons to escape the
residual nuclei is close to unity, $P_{esc}\sim 1$.

As we will show in Sec.~\ref{comparison}, the above-mentioned 
simplifications do not lead to noticeable disagreements with  
experimental data when the removal of one, two or three nucleons is
considered. It means that either above mentioned physical effects are 
negligible, or they  
compensate each other in peripheral nuclear collisions with removal of   
only few nucleons.
However, the predictions of the present model for more 
central collisions with removal of many nucleons 
should be taken with care.

This simple abrasion model can be easily extended to the case of mutual
dissociation. The cross section for the removal of $n_1$ neutrons and
$z_1$ protons from the projectile $(N_1,Z_1)$ {\em simultaneously}
with the removal of $n_2$ neutrons and $z_2$ protons from the
target $(N_2,Z_2)$, ($N_2=A_2-Z_2)$ may be written as:
\begin{equation}
\sigma_{m}^{nuc}(n_1,z_1\mid n_2,z_2)=
\frac{ {Z_1\choose{z_1}} {N_1\choose{n_1}} } 
{{A_1\choose{a}}}\times \sigma^{nuc}(a) \times
\frac{ {Z_2\choose{z_2}} {N_2\choose{n_2}} } 
{{A_2\choose{a}}}.
\end{equation}
Since the number of nucleon-nucleon collisions in such a  
process is assumed to be equal to $a$, the condition 
$z_1+n_1=a=z_2+n_2$ holds.

Using this condition for the process of mutual dissociation with
given numbers of neutrons, $n_1$ and $n_2$, removed from the
projectile and target nuclei, respectively, one has finally:
\begin{equation}
\sigma_m^{nuc}(n_1\mid n_2)=\sum_{z_1}\sigma^{nuc}(n_1,z_1)
\frac{ {Z_2\choose{n_1+z_1-n_2}} {N_2\choose{n_2}} } 
{{Z_2+N_2\choose{z_1+n_1}}}.
\end{equation}   
Here the cross section for single abrasion process, 
$\sigma^{nuc}(n_1,z_1)$,  given by 
Eqs.~(\ref{abr_main}) and (\ref{hyper}) was used.

\section{Input data for heavy ion dissociation calculations}

As shown in Sec.~\ref{EMtheory}, the photonuclear cross sections 
are used as input data in calculations of 
the electromagnetic dissociation cross sections.  
This is verified by the
coherent nature of the photon emission by the collision partner as a 
whole. Since these photons represent the Lorentz-boosted Coulomb
fields of original nuclei, their virtuality is very small, 
$Q^2\leq 1/R^2$, i.e. these photons are almost real. 
Therefore, one can use the 
photonuclear reaction data obtained in experiments with real 
monoenergetic photons
and apply theoretical models describing such photonuclear reactions.

The accuracy of the mutual dissociation cross section calculation
depends heavily on the quality of the data and parameters used as input.
As we found, for example, the mutual dissociation 
cross section is more sensitive to the proper choice of the critical 
impact parameter, $b_c$, and to the input photonuclear cross sections 
than the single dissociation cross section. 
The input data and model parameters are discussed in detail in the
following sections.

\subsection{Photoneutron cross sections measured in 
experiments with real photons}\label{photo_data}

Over the years, the photoneutron cross sections for different nuclei have been
measured with monoenergetic photons at  
Saclay~\cite{Veyssiere,Lepretre,Lepretre3} and Livermore~\cite{Harvey}.  
Data on different cross sections obtained in these and other 
laboratories were collected in compilations of 
Refs.~\cite{Berman-Fultz,Dietrich}. 

\begin{figure}[hp]  
\begin{centering}
\epsfxsize=0.85\textwidth
\epsffile{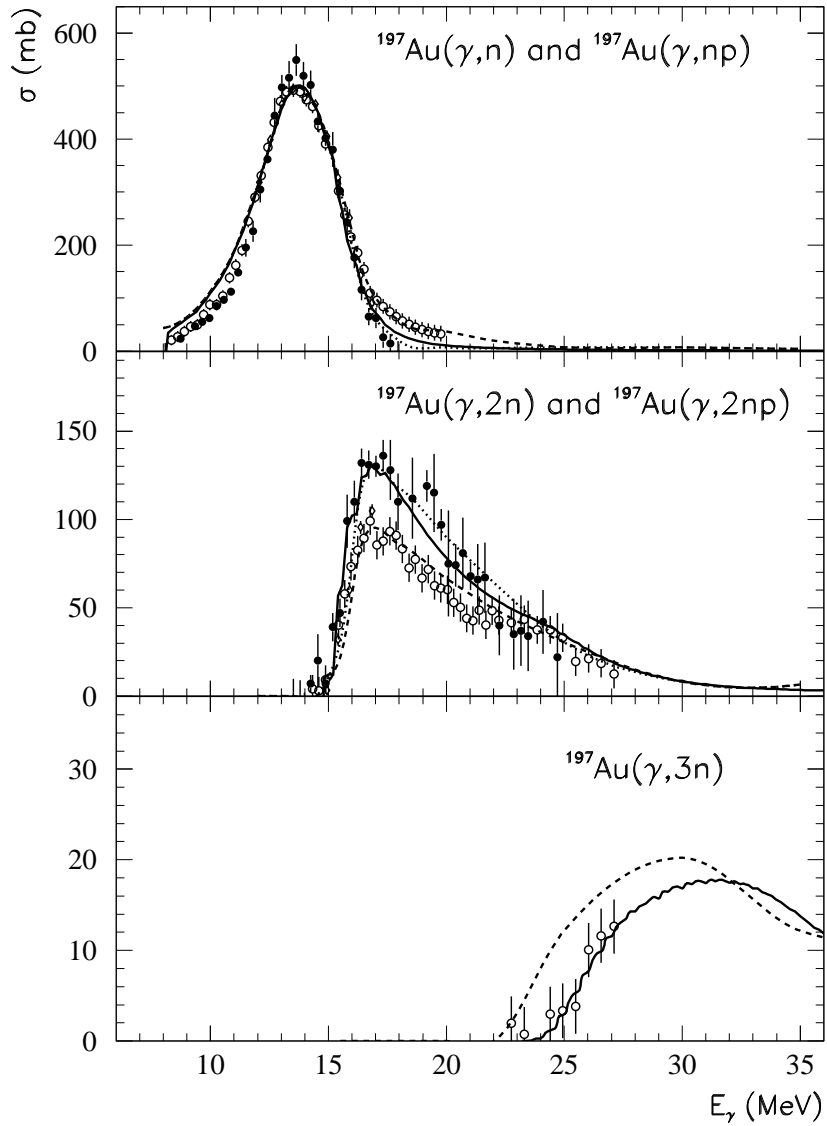}
\caption{Photoneutron cross sections for gold. Open and closed circles are,
respectively, 
Saclay~\protect\cite{Veyssiere} and Livermore~\protect\cite{Harvey} data
re-scaled according to 
Ref.~\protect\cite{Berman}. GNASH code results are presented 
by solid line.  RELDIS results are given by dashed and 
dotted lines for variants with and without inclusion of the
direct $1n$ emission, respectively.}
\label{fig:Au}
\end{centering}
\end{figure}

\begin{figure}[hp]  
\begin{centering}
\epsfxsize=0.85\textwidth
\epsffile{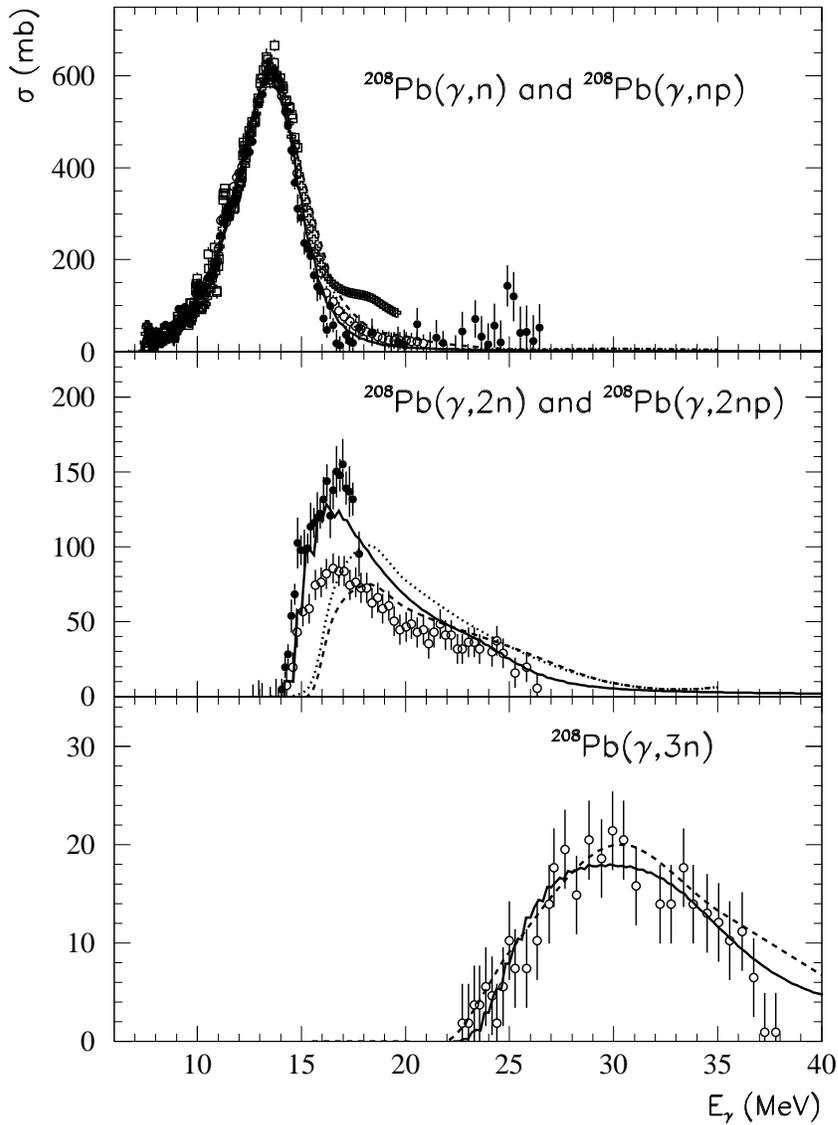}
\caption{Photoneutron cross sections for lead.
Open and closed circles are, respectively,   
Saclay~\protect\cite{Veyssiere} and Livermore~\protect\cite{Harvey} data 
re-scaled according to Ref.~\protect\cite{Berman}. 
Crosses - Saratov data~\protect\cite{SNBeljaev}, squares - 
Moscow evaluated data~\protect\cite{VVVarlamov}. 
Other notations are the same as in Fig.~\protect\ref{fig:Au}.}
\label{fig:Pb}
\end{centering}
\end{figure}

Concerning the nuclei of interest, $^{197}{\rm Au}$ and $^{208}{\rm Pb}$,
the detailed data were obtained mainly for $(\gamma ,n)$ and $(\gamma ,2n)$
reactions, while less detailed data exist for $(\gamma ,3n)$ and
$(\gamma ,4n)$ reactions, see Refs.~\cite{Veyssiere,Harvey}. 
The measurements were performed in the photon energy region 
$6\leq E_\gamma\leq 35$ MeV, where the excitation of giant resonances plays a
dominant role, see Figs.~\ref{fig:Au} and
~\ref{fig:Pb}. At such energies, the emission of charged particles 
$(p,d,^3{\rm He},^4{\rm He})$ is suppressed by a high Coulomb barrier 
of heavy nuclei. Therefore, the sum of partial cross sections for all 
neutron multiplicities:
$\sigma (\gamma ,n)+\sigma (\gamma ,2n)+
\sigma (\gamma ,3n)+\sigma (\gamma ,4n)$
nearly coincides with the total photoabsorption cross section. 
Each of the inclusive cross sections $\sigma (\gamma, in)$ includes a 
small contribution
from the channels with charged particles, $(\gamma ,in\, p)$, 
$(\gamma ,in\, 2p)$,... 
At the same time such channels as $(\gamma ,p)$, $(\gamma ,2p)$ 
were neglected at all. 
According to Ref.~\cite{Lepretre}, this leads to a small
systematic error $\sim 3-5\%$ in the total photoabsorption cross section
measured at low energies via the neutron detection.

Above the giant resonance region, at $35\leq E_\gamma\leq 140$ MeV, 
the quasideuteron
mechanism of photon absorption dominates. 
Only average characteristics of photon
absorption by $^{208}{\rm Pb}$  were measured in 
Saclay~\cite{Lepretre,Lepretre3} in this energy region. Neutron yields,
$\sum_{i\geq 1} i\sigma(\gamma,in)$, and the cross sections for emission of at 
least $j$ neutrons, $\sum_{i\geq j} i\sigma(\gamma,in)$, were obtained 
in addition to the mean value and the width of the neutron multiplicity
distribution.

To the best of our knowledge there are no direct measurements of neutron
emission cross sections and multiplicities at $E_\gamma > 30$ MeV in 
photoabsorption on $^{197}{\rm Au}$. The only attempt to deduce 
the average photoneutron multiplicities from the experimentally 
obtained average excitation 
energies has been made at 
$160\leq E_\gamma\leq 250$ MeV  in a model-dependent way~\cite{Arruda}.

For $(\gamma ,n)$ channel the overall agreement between 
Livermore and Saclay data is good.
Some inconsistency exists only in the giant resonance 
peak height ($\sim 3\%$ for $^{197}{\rm Au}$ and 
$\sim 20\%$ for $^{208}{\rm Pb}$)
and on the right side of the peak, where $(\gamma ,n)$ and  
$(\gamma ,2n)$  process compete with each other, 
see Figs.~\ref{fig:Au} and \ref{fig:Pb}. 
Large discrepancies are present in $(\gamma ,2n)$ cross sections measured 
in different experiments, both in shape and normalization, 
up to $\sim 50\%$ for $^{208}{\rm Pb}$ target.

Several attempts of data evaluation and re-measurements have been made 
to reduce these discrepancies. Based on the observation that the 
total photoneutron yields, 
$\sigma (\gamma ,n)+2\sigma (\gamma ,2n)+3\sigma (\gamma ,3n)$, obtained
in Livermore and Saclay experiments agree well, an explanation
for the discrepancy  
was put forward in Ref.~\cite{Wolynec}. It was attributed to the
different neutron multiplicity sorting procedures 
adopted in different laboratories. 
As was concluded in Ref.~\cite{Wolynec},
the neutron multiplicity sorting procedure adopted at Saclay 
was not correct since some of the $(\gamma,2n)$ events were interpreted 
as pairs of $(\gamma ,n)$ events.

In 1987 new measurements were made in Livermore~\cite{Berman}, 
where it was
found that the previously reported Livermore~\cite{Harvey} and 
Saclay~\cite{Veyssiere} results have to be rescaled.
As was recommended in Ref.~\cite{Berman}, the Saclay data of 
Ref.~\cite{Veyssiere}  for both $^{197}{\rm Au}$ and
$^{208}{\rm Pb}$ nuclei 
have to be used with the correction factor of 0.93.
We follow this prescription in using photonuclear 
cross section data in our RELDIS code. However, such a correction is
not widely accepted and the authors of Refs.~\cite{Weneser,White,Baltz2}
use uncorrected Saclay data. 

Some new $(\gamma ,2n)$ cross section data were obtained in 
Ref.~\cite{Berman} for $^{197}{\rm Au}$ and $^{208}{\rm Pb}$. 
For $^{197}{\rm Au}$ nucleus, $(\gamma ,2n)$ data are nearly 
the same as the Saclay $(\gamma ,2n)$ data of Ref.~\cite{Berman} 
and the conclusion of 
Ref.~\cite{Wolynec} concerning the neutron multiplicity sorting
procedure seems to be not confirmed. Unfortunately, the recent data of
Ref.~\cite{Berman} are available only up to a few MeV above the
$(\gamma ,2n)$ threshold and the findings of Ref.~\cite{Wolynec}
can not be completely ruled out. 

One of the most recent measurements of $(\gamma ,n)$ cross section for
photoabsorption on $^{208}{\rm Pb}$ was performed in Russia at
the Saratov University~\cite{SNBeljaev}. 
A fine structure of the low-energy 
wing of the Giant Dipole Resonance (GDR) has been investigated 
in detail. The photoneutron cross sections were obtained from the photoneutron
yield curves by means of the statistical regularisation method.

An evaluation of $(\gamma, n)$ cross section for $^{208}{\rm Pb}$ has been made
at the Moscow State University~\cite{VVVarlamov} by applying a 
statistical reduction method. 
Because of systematic uncertainties in calibration and normalisation,
the general characteristics of the measured $(\gamma ,n)$ cross section 
(the energy integrated cross sections, weighted-mean values) are different
in different measurements~\cite{Veyssiere,Lepretre,Harvey}.
In the reduction method of Ref.~\cite{VVVarlamov} 
the renormalisation corrections were introduced for both the energy 
and the cross section scales in order to obtain the best agreement between the
general characteristics of  $(\gamma ,n)$ cross section measured 
in different experiments.    

The data obtained in Refs.~\cite{SNBeljaev,VVVarlamov} 
for $(\gamma ,n)$ reaction on $^{208}{\rm Pb}$ target are also plotted
in Fig.~\ref{fig:Pb}. Good agreement with re-scaled Saclay data of 
Ref.~\cite{Veyssiere} is found up to $(\gamma ,2n)$ threshold. Unfortunately,
the $(\gamma ,2n)$ reaction was beyond the scope of investigations in 
Refs.~\cite{SNBeljaev,VVVarlamov}.

It is evident from our consideration that the calculations of electromagnetic
dissociation of ultrarelativistic $^{197}{\rm Au}$ and $^{208}{\rm Pb}$
nuclei can not be based exclusively on the photoneutron cross sections
measured in experiments with real photons. 
Additional information on photoneutron
cross sections for the whole energy domain of equivalent photons
$(E_{min}\leq E_{\gamma}\leq E_{max})$ can be obtained by using theoretical
models of photoabsorption. This is particularly indispensable for dissociation
channels with emission of many neutrons $(\geq 3)$ and charged particles,
$p,d,t,\alpha$,...

\subsection{Evaluation of photoneutron cross sections 
by using the GNASH code}

One of the two photonuclear reaction models used in the present
work is the GNASH code~\cite{Gandini}. It is very precise in describing
low energy neutron emission data~\cite{Chadwick}, however, it can be used 
only up to the pion production threshold, at $E_\gamma\leq 140$ MeV.
Within this model the photoabsorption process is 
modelled through the excitation of the giant dipole 
resonance (GDR) at lower
energies and the quasideuteron (QD) mechanism at higher energies.

The photoabsorption cross section in the whole energy range from the threshold
for neutron emission up to 140 MeV is thus written in the form~\cite{Chadwick}:
\begin{equation}
\sigma_{A}(E_{\gamma})\,=\,\sigma_{GDR}(E_{\gamma})+\sigma_{QD}(E_{\gamma})\,,
\label{Sabs}
\end{equation}
where $\sigma_{GDR}$ is given by a Lorentzian curve with parameters 
taken from GDR systematics~\cite{Dietrich} and corrected
according to Ref.~\cite{Berman}. 
The latter term, $\sigma_{QD}$, is related by a
Levinger-type model to the experimental deuteron photodisintegration cross
section, $\sigma_d$~\cite{Chadwick2}:
\begin{equation}
\sigma_{QD}(E_{\gamma})\,=\,L {NZ \over A} \sigma_d(E_{\gamma}) F(E_{\gamma})\,,
\label{SQD}
\end{equation}
where $N$, $Z$ and $A$ are, respectively, the neutron, proton and mass
number of the corresponding nucleus.
The Levinger parameter, $L$, is equal to 6.5, and $F(E_{\gamma})$
is a Pauli-blocking factor, which reduces the free deuteron cross section,
$\sigma_d(E_{\gamma})$, by taking into account the Pauli 
blocking of the excited
neutron and proton in the nuclear medium. In Ref.~\cite{Chadwick2}, 
$F$ was derived in the
form of a multidimensional integral, approximated in the energy range 20--140 MeV
by a polynomial expression:
\begin{eqnarray}
F(E_{\gamma}) & = & 8.3714 \times 10^{-2} - 9.8343 \times 10^{-3} E_{\gamma}\nonumber \\
              &   &  + 4.1222 \times 10^{-4} E_{\gamma}^2 \label{ff}   \\   
              &   & -3.4762 \times 10^{-6} E_{\gamma}^3 + 9.3537 \times 10^{-9}
                    E_{\gamma}^4 \, . \nonumber 
\end{eqnarray}
and by an exponential one outside the considered energy range:
$$ F(E_{\gamma})\,=\, \left\{
   \begin{array}{l}
   \exp(-73.3/E_{\gamma}), \quad E_{\gamma}\, <\, 20 \, \,{\rm MeV}\\
   \exp(+24.2/E_{\gamma}), \quad E_{\gamma}\, >\, 140\, \,{\rm MeV} 
   \end{array}
    \right. $$
Thus, $F(E_{\gamma})$ tends to zero if $E_{\gamma}$ goes to zero, 
and to unity if $E_{\gamma}$ goes to infinity and is continuous 
with Eq.(\ref{ff}) at 20 and 140 MeV~\cite{Chadwick}.

Finally, the experimental deuteron photodisintegration cross 
section is given by a simple parameterisation:
\begin{equation}
\sigma_d(E_{\gamma})\,=\,61.2 (E_{\gamma}-2.224)^{3/2}/E_{\gamma}^3\,,
\label{sd}
\end{equation}
where $E_{\gamma}$ is expressed in MeV, as in the previous 
formulae, and $\sigma_d$ in mb.

Due to the correlation between intranuclear nucleons in the 
absorption on a quasideuteron pair, the initial particle-hole
configuration is assumed to be $2p1h$ rather than $2p2h$, 
see Ref.~\cite{Chadwick}. In the GNASH code the
initial interaction characterized by the total cross section
of Eqs.~(\ref{Sabs})-(\ref{ff}) is followed by the preequilibrium
emission of fast nucleons described by the exciton 
model~\cite{Gandini,Chadwick}. Finally, when the nuclear system comes
to equilibrium, sequential evaporation of particles is considered
within the Hauser-Feshbach formalism~\cite{Chadwick}.

GNASH code results for $(\gamma ,n)$, $(\gamma ,2n)$ and $(\gamma ,3n)$ 
cross sections are shown in Figs.~\ref{fig:Au} and~\ref{fig:Pb} for
$^{197}{\rm Au}$ and $^{208}{\rm Pb}$ nuclei, respectively. 
Calculations describe
$(\gamma ,n)$ and  $(\gamma ,3n)$ data very well. Taking into account 
existing disagreements between the results of different measurements of
$(\gamma ,2n)$ cross sections, one can conclude that the GNASH 
results fall in between the Saclay~\cite{Veyssiere} and Livermore
data~\cite{Harvey} for $^{208}{\rm Pb}$,  and very close to Livermore data
for $^{197}{\rm Au}$, that seems to be satisfactory for both cases.
Therefore, one can use in Eq.~(\ref{SMFULL}) the 
photonuclear cross sections, $\sigma_A(E_\gamma)$, and
branching ratios, $f(E_\gamma, i)$, calculated by the GNASH code to 
estimate the mutual electromagnetic dissociation
cross section $\sigma_{m}^{ED}(i\mid j)$. The influence of a constraint,
$E_\gamma <E_{max}=140$ MeV, will be discussed in 
Sec.~\ref{LHC}.

\subsection{Neutron emission simulated by cascade and evaporation codes}

Branching ratios for neutron emission in photonuclear reactions,
$f(E_\gamma ,i)$, can be calculated by means of the extended 
cascade-eva\-pora\-tion-fis\-sion-multi\-frag\-men\-tation model 
of photonuclear
reactions~\cite{Iljinov} in the whole range of equivalent photon energies.
Some details of the calculation method as well as numerous comparisons 
with experimental data used for the model verification were given in
Refs.~\cite{Pshenichnov2,Pshenichnov,Iljinov}. Here we describe 
only the general
calculation scheme along with the modifications and advancements
made in the model since the time when the 
works~\cite{Pshenichnov2,Pshenichnov,Iljinov} have been published. 

In the RELDIS model the values of the total photoabsorption cross section
to be used in Eq.~(\ref{SMFULL}) are taken from corresponding
approximations of experimental data. In the GDR region the Lorentz
curve fits were used for this purpose with parameters from 
Ref.~\cite{Berman-Fultz} corrected according to the prescription of 
Ref.~\cite{Berman}, as described in Sec.~\ref{photo_data}.
Above the GDR region, where the quasideuteron absorption comes into play,
the total cross section is taken from the quasideuteron model of
Ref.~\cite{Lepretre}:
\begin{equation}
\sigma_{QD}(E_\gamma)=k{NZ\over A}\sigma^{exch}_d(E_\gamma).
\end{equation}
Here $\sigma^{exch}_d$ is the meson exchange part of the cross section
for the deuteron photodisintegration, $\gamma d\rightarrow np$, and $k$ is
an empirical constant~\cite{Lepretre}.

Above the pion production threshold, at $E_\gamma\geq 140$ MeV, a 
universal behaviour $\sigma_A(E_\gamma)\propto A$ is observed
(see Ref.~\cite{Valeria} for the latest experimental data). This
means that the total photoabsorption cross section per bound nucleon
$\sigma_A(E_\gamma)/A$ has almost the same energy dependence for 
light, medium-weight and heavy nuclei, C, Al, Cu, Sn, Pb, 
at least up to $E_\gamma \sim 3$ GeV. Therefore, having the data for one
nucleus one can calculate the cross section for other nuclei.
However, in this energy region the universal curve $\sigma_A(E_\gamma)/A$ 
is very different from the values extrapolated from
the cross sections on free nucleons,
$(Z\sigma_{\gamma p}+N\sigma_{\gamma n})/A$, which are deduced from 
proton and deuteron data~\cite{Valeria}. At $E_\gamma > 3$ GeV 
the universal behaviour breaks down, and the ratio $\sigma_A(E_\gamma)/A$ 
for lead is  20-25\% lower than for 
carbon~\cite{Engel-Roesler,Bianchi} due to the nuclear shadowing effect.
In order to approximate 
the total photonuclear cross sections at $E_\gamma > 3$ GeV
we used recent results obtained within the framework of 
the Glauber-Gribov scattering theory and the
Generalized Vector Dominance model~\cite{Engel-Roesler,Bianchi}.
Such calculations describe well the general trend of
experimental data obtained for high energy photon absorption, 
although the data have very large uncertainties at $E_{\gamma} > 10$ GeV.

By comparing Tab.II of Ref.~\cite{Pshenichnov2} and Tab.~\ref{T1}
of the present paper one can find minor differences in the total 
ED cross sections due to using different parameterisations 
of $\sigma_A(E_\gamma)/A$ at $E_\gamma > 3$ GeV. 
Compared with the total ED cross sections, the single 
or double neutron emission cross sections are even less  
affected by the choice of the parameterisation.

The RELDIS code performs the Monte Carlo simulation of the mutual
dissociation process according to the following steps. First, a pair of
the energies $E_1$ and $E_2$ of the photons exchanged between 
the colliding nuclei is
selected according to the spectral function ${\cal N}_m(E_1,E_2)$.
Second,  the photoabsorption process is generated 
in both nuclei leading to the formation of 
excited residual nuclei. Third, the de-excitation of both of the thermalized
residual nuclei is simulated according to the statistical
evaporation-fission-multifragmentation model, (the SMM model)~\cite{JPB}.

The evaporation of neutrons from excited residual compound-like 
nucleus is the main process responsible 
for the $(\gamma, n)$, $(\gamma, 2n)$, $(\gamma, 3n)$
channels of photoabsorption. The quality of description of such channels
is very important for precise calculations of neutron emission in the mutual
dissociation process. In the present paper the standard Weisskopf evaporation 
scheme is used~\cite{JPB} with several modifications taking into account 
the microscopic effects of nuclear structure in the nuclear mass 
and level density formulae. Such effects reveal themselves in 
the noticeable difference, up to $\sim 10-15$ MeV for heavy closed-shell 
nuclei, between the values of the nuclear mass measured in experiments 
and those predicted by the macroscopic liquid-drop model. 

Moreover, this difference in mass, the so-called shell correction, and 
the level density parameter are strongly correlated. 
For closed-shell nuclei the actual values of the level density 
parameter are essentially lower than the average values 
expressed as $A/8-A/10$ MeV$^{-1}$, 
and these values depend noticeably on the excitation energy.     
Proper accounting for these effects, as well as pairing effects, 
is important at low excitations, $E^\star \sim 10$ MeV, 
i.e. in the region where $1n$ and $2n$ photoemission processes occur.

The above-mentioned shell effects are most pronounced at low excitation
energies, but almost disappear at $E^\star >50$ MeV, see 
Ref.~\cite{Gaimard} for details. Several phenomenological systematics
of the level density parameter were proposed to account for such behaviour, 
Refs.~\cite{Ignatyuk75,Ignatyuk79,ILJ-MEB}.
Our calculations are based on results of Ref.~\cite{ILJ-MEB}
where all existing data on the level densities, decay widths and
lifetimes of excited nuclei have been analyzed in the framework of the
statistical model.  

However, the creation and subsequent decay of an excited compound nucleus
formed after the photoabsorption in GDR region is not the only process
responsible for the neutron emission. Indeed, a giant resonance is  
a coherent superposition of (one-particle-one-hole) $1p1h$ excitations.
A particle or a hole can interact with another nucleon and create a $2p2h$
state. Further spreading to $3p3h$ states etc. finally leads to a 
statistically equilibrated system, the compound nucleus. Instead of such
evolution to equilibrium, a collective $1p1h$ state can decay directly
by the emission of one nucleon leading to a low-lying hole state in 
the residual nucleus, see among others Ref.~\cite{VandenBerg}. 
After such direct $1n$ emission, the emission of a second neutron
is generally impossible, even though the initial photon energy 
exceeds the $2n$ emission threshold. In such a way $(\gamma ,2n)$
channel is suppressed in comparison with the pure statistical decay. 

Although the GDR state in $^{208}{\rm Pb}$ nucleus decays mainly statistically,
the existence of direct neutron emission has been clearly 
demonstrated in Ref.~\cite{Alarcon}. The process where a fast nucleon is
emitted and the final state $^{207}{\rm Pb}$ nucleus is left with low
excitation energy $\leq 3$ MeV was identified in this 
experiment~\cite{Alarcon}. An evidence of direct neutron emission  in
photoabsorption on ${\rm Au}$ and ${\rm Pb}$ nuclei was given in 
Ref.~\cite{Veyssiere} based on the analysis of competition 
between $1n$ and $2n$ emission channels. This was an independent 
confirmation of the findings of earlier 
works~\cite{Askew} and \cite{Tagliabue} devoted to the measurements 
of the neutron spectra in photoabsorption on the same nuclei.
An excess of fast neutrons (kinetic energy $\geq 4$ MeV) with 
respect to the predictions
of the statistical evaporation model has been demonstrated and attributed to
the direct emission.

In Ref.~\cite{VandenBerg} a non-statistical contribution in 
excited $^{208}{\rm Pb}$ nucleus with 
$10 < E^\star < 30 $ MeV was successfully extracted. 
Out of this region the non-statistical 
contribution was found to be negligible. In our calculations we 
used the total fractions of the non-statistical neutron emission
from ${\rm Au}$ and ${\rm Pb}$ nuclei as $P^{dir}_n=0.31$  and 0.26,
respectively, evaluated from experimental data in Ref.~\cite{Veyssiere}.
Such values are in line with modern theoretical 
expectations~\cite{Chekomazov} that the ratio of intensities of the
direct and statistical neutron emission from photoexcited 
GDR in $^{208}{\rm Pb}$  
nucleus is about $\sim 0.1$. In the RELDIS code the emission angles
$\Theta$ of non-statistical fast photoneutrons are generated according
to the approximation $W(\Theta)=A+B\sin^2{\Theta}$ which is found 
in Ref.~\cite{Tagliabue}. We assumed that the direct $1n$ emission 
takes place at $7\leq E^\star \leq 22$ MeV. 
 
Since the adopted $P^{dir}_n$ values have some uncertainties, we have
investigated the sensitivity of results to these values. 
A part of calculations was made with $P^{dir}_n=0$, i.e. without
accounting for direct emission, see Figs.~\ref{fig:Au}, \ref{fig:Pb}
and Tab.~\ref{THill}.
As shown in Fig.~\ref{fig:Au}, the 
$(\gamma, 2n)$ cross sections on gold calculated by the RELDIS
code with $P_n^{dir}=0.31$ are very close to 
Saclay measurements~\cite{Veyssiere}, while Livermore 
results~\cite{Harvey} are better
described with  $P_n^{dir}=0$.
Therefore the difference in
calculation results obtained with $P_n^{dir}=0$ and $P_n^{dir}=0.31$ 
reflects the level of experimental uncertainties.

\subsection{Choice of cutoff parameter and nuclear density
distributions}\label{parameters}

Since the nucleon-nucleon interaction has an isovector component, 
the interference of nuclear and electromagnetic amplitudes cannot
be excluded. Such interference term was considered in Ref.~\cite{BenCook}
and found to be small. Even for $^{197}{\rm Au}$ nucleus colliding
with heavy targets the interference
correction to the single neutron removal cross section 
was found to be less than 0.5--0.6 \% of the 
corresponding nuclear or electromagnetic contributions.
Following this result, nuclear and electromagnetic parts of the 
dissociation cross section may be safely treated separately. In other
words, one can add probabilities                   
instead of coherent summation of amplitudes.    

Let us consider the way how the probabilities of the nuclear and
electromagnetic contributions should be added to obtain the total
dissociation probability. At grazing impact parameters
relativistic nuclei are partly transparent 
to each other. Hard
$NN$ collisions may be absent at all in the case of a peripheral 
event with a weak overlap of diffuse nuclear surfaces, 
while the electromagnetic interaction
may take place in this event leading to the electromagnetic 
dissociation. 
Generally, at a grazing collision either the nuclear, or 
electromagnetic interaction, or even both of them may occur. 
As an example of the latter case,
a single neutron-neutron collision in the participant 
zone may lead to the neutron removal, while a photon may be emitted 
and absorbed by charged spectators in the same event.  
 
Therefore, in a detailed theoretical model a smooth transition from purely 
nuclear collisions at $b\ll R_1+R_2$ to electromagnetic 
collisions at $b\gg R_1+R_2$ should take place. Such kind of transition 
was considered in a ``soft-sphere'' model of Ref.~\cite{Aumann}. 
A similar approach was adopted in Ref.~\cite{Baltz2}, where the cross 
section for at least one type of dissociation, either nuclear, 
or electromagnetic, or both, was written as: 
\begin{equation}
\sigma = 2\pi \int^{\infty}_{0} bdb 
\biggl({\cal P}^{nuc}(b)+{\cal P}^{ED}(b)-{\cal P}^{nuc}(b){\cal P}^{ED}(b)
\biggr),
\end{equation}
where ${\cal P}^{nuc}(b)$ and ${\cal P}^{ED}(b)$ are, respectively, 
the probabilities
of the nuclear and electromagnetic dissociation at given impact
parameter $b$. Putting explicitly the integration limits for each 
term one obtains:
\begin{equation}
\sigma = 2\pi \int^{b^{nuc}_{c}}_{0} bdb {\cal P}^{nuc}(b)+
2\pi \int^{\infty}_{b^{ED}_{c}} bdb {\cal P}^{ED}(b)
- 2\pi \int^{b^{nuc}_{c}}_{b^{ED}_c} bdb {\cal P}^{nuc}(b){\cal P}^{ED}(b).
\label{FULLP}
\end{equation}
Here the impact parameter cutoff values, $b^{nuc}_{c}$ and  
$b^{ED}_{c}$, were used for the nuclear and electromagnetic interactions,
respectively. However, due to several reasons a more simple expression is  
widely used~\cite{Krauss,BaurHen}:
\begin{equation}
\sigma = \sigma^{ED}+\sigma^{nuc} =
2\pi\int^{b_{c}}_{0} bdb {\cal P}^{nuc}(b)+
2\pi\int^{\infty}_{b_{c}} bdb {\cal P}^{ED}(b),
\label{TRUNCATEDP}
\end{equation}
where a single cutoff parameter, $b_c$, was chosen as: 
$b^{ED}_c<b_c<b^{nuc}_c$. The first reason is that
with the latter condition one can effectively reduce the 
first and second terms of Eq.~(\ref{FULLP}) without subtracting
the third nuclear-plus-electromagnetic term.
Numerical results based on Eqs.~(\ref{FULLP})
and (\ref{TRUNCATEDP}) become similar as
it was found for ``sharp-cutoff'' and ``soft-spheres''
models of Ref.~\cite{Aumann}.  Second, for heavy nuclei
the difference between reasonable values of 
$b^{ED}_c$, $b^{nuc}_c$ and $b_c$ turns out to be less than 1 fm.
As a result, the last nuclear-plus-electromagnetic term of 
Eq.~(\ref{FULLP}) turns out to be small. 
Third, with Eq.~(\ref{TRUNCATEDP}) one can study the nuclear 
and electromagnetic contributions separately.  
Therefore, independent parameterisations may be found in experiments 
for the nuclear and electromagnetic parts, $\sigma^{nuc}$ and
$\sigma^{ED}$. This is useful for studying nuclear and electromagnetic
dissociation at ultrarelativistic  
colliders, where the products of nuclear and 
electromagnetic interactions populate very different rapidity regions,
namely the central rapidity region and close to the beam rapidity,
respectively.  

In the present paper the traditional 
form given by Eq.~(\ref{TRUNCATEDP}) is adopted with a common 
impact parameter cutoff, $b_c$, for nuclear and 
electromagnetic contributions. 
At relativistic energies, according to the widely 
used BCV parameterisation of Ref.~\cite{BenCook}, $b_c$ 
is estimated as:
\begin{equation}
b_c=R_{BCV}\bigl( A_1^{1/3}+A_2^{1/3}-X_{BCV}(A_1^{-1/3}+A_2^{-1/3})\bigr).
\label{BCV}
\end{equation}
The values  $R_{BCV}=1.34$ fm and $X_{BCV}=0.75$ were found from a fit to
Glauber-type calculations of the total nuclear 
reaction cross sections~\cite{BenCook}.   

The evidences in favour of the $b_c$ choice according to the BCV 
parameterisation were given in Refs.~\cite{Norbury},\cite{Grunschloss},
which we mention among others. As argued in 
Ref.~\cite{Grunschloss}, using the BCV parameterisation  
one can perfectly describe experimental data on fragment angular
distributions which are very sensitive to $b_c$.

For calculations within the abrasion model we used the following
values for the total nucleon-nucleon cross section given in 
Ref.~\cite{RPP}: 
$\sigma_{NN}=40$, 50 and 90 mb at SPS, RHIC and LHC energies 
respectively. 
Some problems are connected with the choice of the 
nuclear density parameters, 
$R_o=r_o\times A^{1/3}$, and $d=0.54$ fm, Sec.~\ref{abrmodel}.
Only the nuclear charge distributions are
measured in electron scattering experiments, while the
neutron densities are available only from calculations. 
We used $r_o=1.14$ fm as an average value between
the proton and neutron distributions 
similar to one used in Ref.~\cite{Baltz2}. 
The total nuclear reaction cross sections calculated with these 
$r_o$ and $d$ values 
within the abrasion model are in good agreement with
the approximation of experimental data found in Ref.~\cite{BenCook}.
 
Numerical results showing the sensitivity of the nuclear and 
mutual electromagnetic dissociation cross sections to the 
variations of the above-discussed parameters are given in Sec.~\ref{LHC}.

\section{Comparison with CERN SPS data}\label{comparison}

\begin{figure} 
\begin{center} 
\vspace{1cm} 
\unitlength=1cm 
\begin{picture}(8.0,2.0) 

\SetWidth{1.0}
\ArrowLine(0,100)(60,100)  
\Text(0,3.9)[]{$A_1$} 
\SetWidth{0.5}  
\GCirc(60,100){4}{1.} 
\SetWidth{1.0}
\ArrowLine(64,100)(120,100)  
\Text(3.1,3.9)[]{$A_1$}   

\SetWidth{0.5}
\Photon(60,97)(60,36){3}{4} 
\Text(1.7,2.2)[]{$E_1$} 

\SetWidth{1.0}
\ArrowLine(0,33)(60,33)  
\Text(0,0.8)[]{$A_2$} 
\SetWidth{0.5}  
\GCirc(60,33){4}{0} 
\SetWidth{2.0}
\ArrowLine(60,33)(120,33)  
\Text(3.1,0.8)[]{$A_2^\star$}
\SetWidth{0.5}  
\GCirc(120,100){4}{1}
\SetWidth{2.0} 
\ArrowLine(120,33)(180,33)  
\Text(5.9,0.8)[]{$A_2^\star$}
\SetWidth{0.5}  
\GCirc(120,33){4}{0} 
  
\Photon(120,97)(120,36){3}{4} 
\Text(3.9,2.2)[]{$E_2$}

\SetWidth{1.0}
\ArrowLine(124,100)(180,100)  
\Text(5.9,3.9)[]{$A_1$}   
 
\end{picture} 
\end{center} 
\caption{ 
Electromagnetic excitation of a single nucleus: second order 
process. Open and closed circles denote elastic and inelastic vertices,
respectively.  
\label{EM2}  
} 
\end{figure}
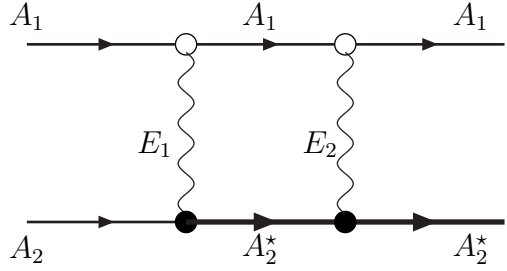

\begin{figure}[ht]
\begin{centering}
\epsfxsize=0.85\textwidth
\epsffile{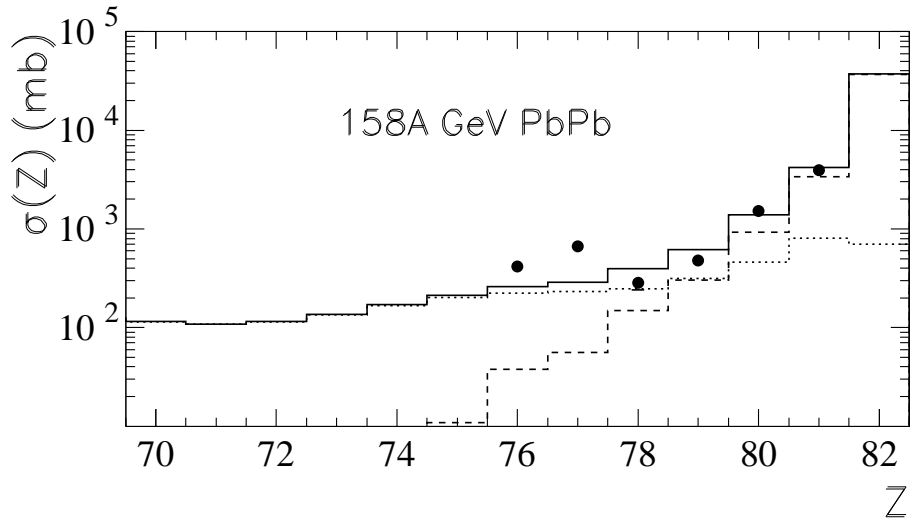}
\caption{Charge changing cross sections of 158 AGeV 
$^{208}{\rm Pb}$ ions on Pb target.  
The dashed- and dotted-line histograms are the RELDIS and 
abrasion model results for electromagnetic and nuclear 
contributions, respectively. The solid line histogram 
presents the sum of both contributions.
Points are experimental data from Ref.~\protect\cite{Dekhissi}.}
\label{fig:zdis}
\end{centering}
\end{figure}

The calculated charge changing cross 
sections of the {\em single} dissociation of 
158A GeV $^{208}{\rm Pb}$ ions are given in Fig.~\ref{fig:zdis}.
The LO mechanism with single photon exchange, Fig.~\ref{EM}, and 
the NLO mechanism with double photon exchange,  Fig.~\ref{EM2}, 
were taken into account. 

Each proton removal process can be accompanied by neutron loss
as well. The calculated cross sections given for $Z=82$ 
correspond to the interaction where only neutrons are emitted.
However, experimental data are absent for this channel. 
As one can see, the electromagnetic contribution dominates for
the processes with removal of one, two and three protons where good
agreement with the experimental data of Ref.~\cite{Dekhissi} is found.
 
Another important check of the model becomes possible with recent
experimental data for Au fragmentation by ultrarelativistic
Pb ions~\cite{Hill-Petridis}. In this case the neutron emission
is investigated directly. The experimental exclusive 
cross sections for emission of one or two neutrons are compared
with theory in Tab.~\ref{THill}. Also in this case the 
contributions from the LO and NLO processes, Figs.~\ref{EM} 
and \ref{EM2}, were taken into account. 

The calculations were made
with and without accounting for direct
neutron emission process, i.e. with $P_n^{dir}=0.31$ and  $P_n^{dir}=0$,
respectively. The results with $P_n^{dir}=0.31$ are in 
better agreement with experiment and this is especially true for
$2n$ emission channel. Therefore, we use this value in further 
calculations.

The predictions of abrasion model are also in good agreement with
data~\cite{Hill-Petridis}. Therefore, our choice of the critical
impact parameter, $b_c$, is justified by such a comparison.
As was stressed in Sec.~\ref{abrmodel}, the interaction of knocked 
out nucleons with spectators and spectator de-excitation process themselves 
were neglected in this version of the abrasion model aimed at
considering only few nucleon removal processes.
Such good agreement indicates that a simple abrasion model
proposed for describing the data at $\sim 1$--10 AGeV can be
used successfully at much higher energies as well.

\section{Mutual dissociation of $^{197}{\rm Au}$  and 
$^{208}{\rm Pb}$ ions at RHIC and LHC}\label{LHC}

On the basis of the successful verifications 
at lower energies described in Sec.~\ref{comparison}, 
the model can now be extrapolated to RHIC and LHC energies.
In a collider, the mutual heavy-ion dissociation process 
takes place at the crossing point of two beams.
Downstream from this point the dissociation products
can be separated by the magnetic field according to their
$Z/A$ ratio.  Protons and nuclear fragments move 
close to the beam trajectories, 
while free neutrons leave the beam pipe after a dipole magnet. 

At RHIC, Zero Degree (very forward) Calorimeters (ZDC) for each 
beam are located after the magnets and they are well
designed for the neutron registration~\cite{Baltz,White,Baltz2}. 
Therefore, in the following we consider mainly semi-inclusive mutual 
neutron emission cross sections, $\sigma_m(i\mid j)$,
where  $i$ and $j$ denote corresponding  
channels, $1nX$, $2nX$, $3nX$,... Besides the emission of a given 
number of neutrons, such dissociation channels contain  
any number of other particles denoted as $X$ or $Y$. 
Protons can be found most often amongst the particles emitted along with 
neutrons. The proton emission rates predicted by the RELDIS code were 
found to be in agreement with the data on Pb dissociation~\cite{Dekhissi},
see Sec.~\ref{comparison}.
Therefore we believe that our model is accurate in estimating 
$\sigma_m(1nX\mid 1nY)$, $\sigma_m(1nX\mid 2nY)$ and 
$\sigma_m(2nX\mid 2nY)$ cross sections. 

If the dissociation channel of one of the collision partners 
in mutual dissociation is not exactly known, one can define 
inclusive mutual dissociation cross sections, $\sigma_m(1nX\mid {\cal D})$, 
$\sigma_m(2nX\mid {\cal D})$, $\sigma_m(3nX\mid {\cal D})$ etc.
In such notations ${\cal D}$ denotes an arbitrary 
dissociation mode of one of the nuclei. 

The cross sections for some specific channels of mutual dissociation 
at RHIC energies are given in Tab.~\ref{TLONLO}. 
Compared to the total cross sections, the relative contributions of NLO 
processes are very different for the cross sections of $1n$, $2n$ and
$3n$ emission. As one can note, $\sigma_m^{ED}(1nX\mid 1nY)$ has a small
NLO correction, $\sim 7$\%, while $\sigma_m^{ED}(3nX\mid D)$ becomes almost
twice as large as its LO value if NLO correction is included. 
This is due to the fact that
the NLO processes shown in Figs.~\ref{EMNLO12} and \ref{EMNLO22} include
nuclear excitation due to double photon absorption and, particularly,
the double GDR excitation process. Since the average GDR energy for Au and Pb
nuclei is about 13--14 MeV, the double GDR excitation introduces, on average, 
26--28 MeV excitation which is above the $3n$ emission threshold.

Therefore, as a rule, $1n$ and $2n$ emission cross sections are less
affected by including of NLO corrections than $3n$ cross sections.      
The double GDR excitation process is a well-known phenomenon  
studied in low and intermediate energy heavy ion collisions, 
see, for example, Ref.~\cite{Bertulani-Ponomarev}. 
However, nothing is known about the multiple photon excitations of heavy ions 
at ultrarelativistic energies, where the key question is to what extent
the distribution of probabilities for multiple excitations follows
the Poisson distribution (the harmonic picture of excitations).

In order to estimate the rates of multiple 
excitations, including those well above the giant resonance 
region, the ratios 
$\sigma_m^{ED}(2nX\mid D)/\sigma_m^{ED}(1nX\mid 1nY)$ and  
$\sigma_m^{ED}(3nX\mid D)/\sigma_m^{ED}(1nX\mid 1nY)$ should be measured at RHIC.
Such measurements do not require preliminary determination of the  
collider luminosity. As it follows from Tab.~\ref{TLONLO}, the above-mentioned
ratios are remarkably reduced if the multiple excitations, and hence
NLO processes, are suppressed in mutual electromagnetic dissociation at RHIC.

There remains still some freedom in choosing several parameters of our model.
To check the sensitivity of our predictions to their variations
we performed the calculations for a reasonable span
of input parameters.  

Tab.~\ref{TPdir} demonstrates the sensitivity of the mutual 
electromagnetic dissociation 
cross sections to the photonuclear cross sections used
as the input. In order to demonstrate such sensitivity
we used two different models to calculate such cross sections, namely
the GNASH code~\cite{Gandini} and the photonuclear reaction 
model implemented in the RELDIS code~\cite{Pshenichnov2,Iljinov} itself.
Additionally, in the latter model we used two different values for
the probability of the direct neutron 
emission in $1n$ channel, $P^{dir}_n$.

Besides the variations of the cross sections for emission of one or 
two neutrons, the variations of a cumulative value, 
the Low Multiplicity Neutron (LMN) emission 
cross section defined as
\begin{eqnarray*}
\sigma_m(LMN)=\sigma_m(1nX\mid 1nY)+\sigma_m(1nX\mid 2nY)+ \\
\sigma_m(2nX\mid 1nY)+\sigma_m(2nX\mid 2nY) 
\end{eqnarray*}
were evaluated for several sub-regions of equivalent photon energies,
$E_\gamma<24$~MeV, $E_\gamma<140$ MeV and for the full range.
In the latter case the NLO processes were taken into account along with the
LO mutual dissociation process. Therefore, 
the importance of NLO corrections can be also estimated from 
Tab.~\ref{TPdir}. 
 
By examining Tab.~\ref{TPdir}, one can draw several conclusions.
First, the  semi-inclusive cross section $\sigma^{ED}_m(1nX\mid 1nY)=437$ mb 
calculated for the photoabsorption in the giant resonance region 
is very close to the exclusive value $\sigma^{ED}_m(1n\mid 1n)=445$ mb 
obtained in Ref.~\cite{Baltz2} with the same condition: $E_\gamma<24$ MeV.
Second, the calculations based on the photonuclear cross sections predicted
by the GHASH code are very close to the RELDIS results for the photon
energy region $E_\gamma<140$ MeV. The difference
between the RELDIS results for $E_\gamma<24$ MeV and $E_\gamma<140$ MeV
is explained by the contribution of the quasideuteron photoabsorption mechanism
to the $1n$ and $2n$ emission.
Third, the calculations which take into account 
the quasideuteron photoabsorption and photoreactions
above the pion production threshold give about 25\% enhancement in  
$\sigma^{ED}_m(1nX\mid 1nY)$
if the whole energy region of equivalent photons is considered and
the NLO corrections are properly taken into account.
At the same time, the cross sections $\sigma^{ED}_m(1nX\mid 2nY)$ and
$\sigma^{ED}_m(2nX\mid 2nY)$ increase up to two and four times, 
respectively, compared with the
values calculated at the GDR region.

The cross sections for these
dissociation channels are large and such channels can be easily 
measured in experiments. Although the LO process of 
photoabsorption in the GDR region gives an important contribution, 
the whole range of the equivalent photon energies 
should be considered. Also the contributions from the NLO processes 
should be taken into account to obtain
the realistic values of the dissociation cross sections.

One more conclusion follows from the results presented in Tab.~\ref{TPdir}.
Calculations with $P^{dir}_n=0$ and $P^{dir}_n=0.31$ give 10-40\%
difference in specific dissociation cross sections, but
the values of the LMN cross section, $\sigma^{ED}_m(LMN)$,
practically coincide.
This cross section is very high, $\sigma^{ED}_m(LMN)\sim 1100$ mb, 
and thus can be used for luminosity monitoring. As was shown above in 
Sec.~\ref{photo_data}, an inevitable systematic
error of $\sim 5\%$ should be assigned to this value due to 
uncertainties in the photoneutron cross sections measured in experiments. 

Tab.~\ref{Tro} shows the sensitivity of the mutual dissociation cross 
section in grazing nuclear collisions to the variations of the $r_o$ 
parameter in the nuclear density distribution, Eq.~(\ref{Fermi}).
The parameters of neutron density distributions are not well 
determined and this table demonstrates possible ambiguities 
in nuclear dissociation cross sections caused by this fact.  
The cross section variations are smaller in Tab.~\ref{Tro} 
compared to Tab.~\ref{TPdir}, about 3-8\%.
A small decrease in the $r_o$ parameter leads to a decrease in
correlated $1n-1n$ emission, but, on the contrary, it leads to an increase in
$1n-2n$ and $2n-2n$ emission. 
However, the LNM cross section, $\sigma^{nuc}_m(LMN)$, turns out to be 
more stable, within $\sim 2$\%  
variations, compared to the cross sections for specific channels.

The sensitivity of the dissociation cross 
section in grazing nuclear collisions to the variations of the 
total nucleon-nucleon cross section is investigated in Tab.~\ref{TNN}.
The variations of the $\sigma_{NN}$ in the range of 40--60 mb 
have only a small effect, within  4\%, 
on the cross sections of the specific neutron emission channels.
Since such variations have different signs, the influence on the 
cumulative value, $\sigma^{nuc}_m(LMN)$, is less noticeable, 
below 2\%. 
  
The cross sections given in Tabs.~\ref{Tro} and \ref{TNN} 
for grazing nuclear collisions were found to be lower compared with 
the electromagnetic dissociation cross sections of Tab.~\ref{TPdir}.
Moreover, it should be stressed, that only a part of the neutrons
in grazing nuclear collisions is emitted at the forward or backward
angles covered by the ZDCs. Therefore, an exact relation between 
nuclear and electromagnetic dissociation channels in each 
heavy-ion experiment should be only obtained by taking into 
account detection acceptances and trigger conditions of the corresponding 
experimental setup.   

Tab.~\ref{Tbc} demonstrates the sensitivity of the mutual dissociation 
cross section in electromagnetic and nuclear
interactions to the impact parameter cutoff, $b_c$. 
This is an important input parameter which has
a noticeable influence on the final result.
By changing this parameter by 5\%, within the range of
14.5--16 fm, one obtains the variations of the electromagnetic
dissociation cross sections within 3--8\%. 
Such variations of $b_c$ shift the point, which delimits the regions of nuclear
and electromagnetic interactions, below or above the domain where the
overlap of diffuse nuclear boundaries takes place. In other words,
assuming first $b_c\approx b^{ED}_c$ and then $b_c\approx b^{nuc}_c$ and
considering the difference in final results,
one can prove the possibility to use Eq.(\ref{TRUNCATEDP}) instead
of Eq.(\ref{FULLP}). 
     
For example, if the cutoff value, $b_c$, becomes lower, 
all the ED cross sections,
$1n-1n$, $1n-2n$, $2n-2n$ and  $\sigma^{ED}_m(LMN)$ become lower. 
The variations of the nuclear cross sections are more noticeable and 
have the opposite trend: such cross sections become higher by
5--15\%.  
Finally, as one can see in Tab.~\ref{Tbc}, $\sigma_m(LMN)$ 
variations for both types of interactions are
weaker than the variations of the corresponding individual cross sections, 
within 3--8\%, 
while the sum $\sigma^{ED}_m(LMN)+\sigma^{nuc}_m(LMN)$ is altered by 1-4\% 
only.

Concluding the investigation of the sensitivity of 
the final results to the model parameters, 
one can note that for AuAu collisions at RHIC
$\sigma^{ED}_m(LMN)=1100$ mb  
and especially $\sigma^{nuc}_m(LMN)=737$ mb values are more stable
with respect to such variations in comparison with the individual
cross sections, $\sigma_m(1nX\mid 1nY)$, $\sigma_m(1nX\mid 2nY)$ 
and others.   

The same tendency was found for PbPb collisions at LHC energies where
such cross sections were found to be $\sigma^{ED}_m(LMN)=1378$ mb
and $\sigma^{nuc}_m(LMN)=755$ mb.
For both RHIC and LHC cases the overall uncertainty of the 
$\sigma_m(LMN)$
calculation method may be estimated at the level of 5--7\%.

\begin{figure}
\begin{centering}
\epsfxsize=0.7\textwidth
\epsffile{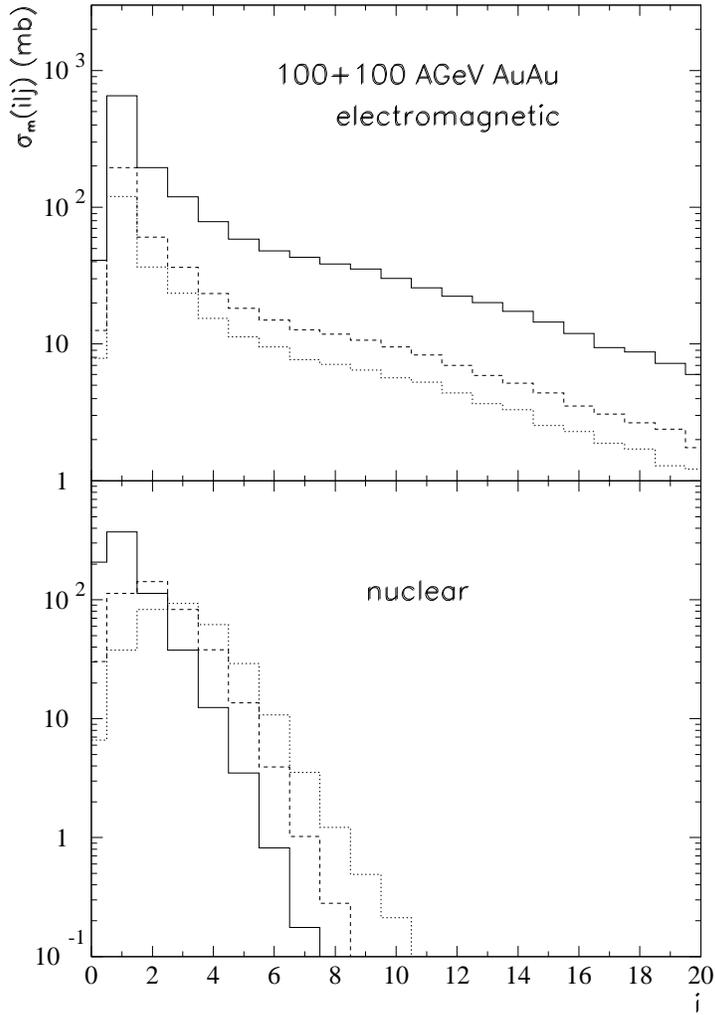}
\caption{Mutual electromagnetic, $\sigma^{ED}_m(i\mid j)$ (top), and
nuclear, $\sigma^{nuc}_m(i\mid j)$ (bottom),  
dissociation cross sections for neutron 
emission in 100+100 AGeV AuAu collisions at RHIC.
The cross section values for $j=1nX,2nX,3nX$ are given by the solid,
dashed and dotted histograms, respectively.}
\label{fig:rhic2d}
\end{centering}
\end{figure}

\begin{figure}
\begin{centering}
\epsfxsize=0.7\textwidth
\epsffile{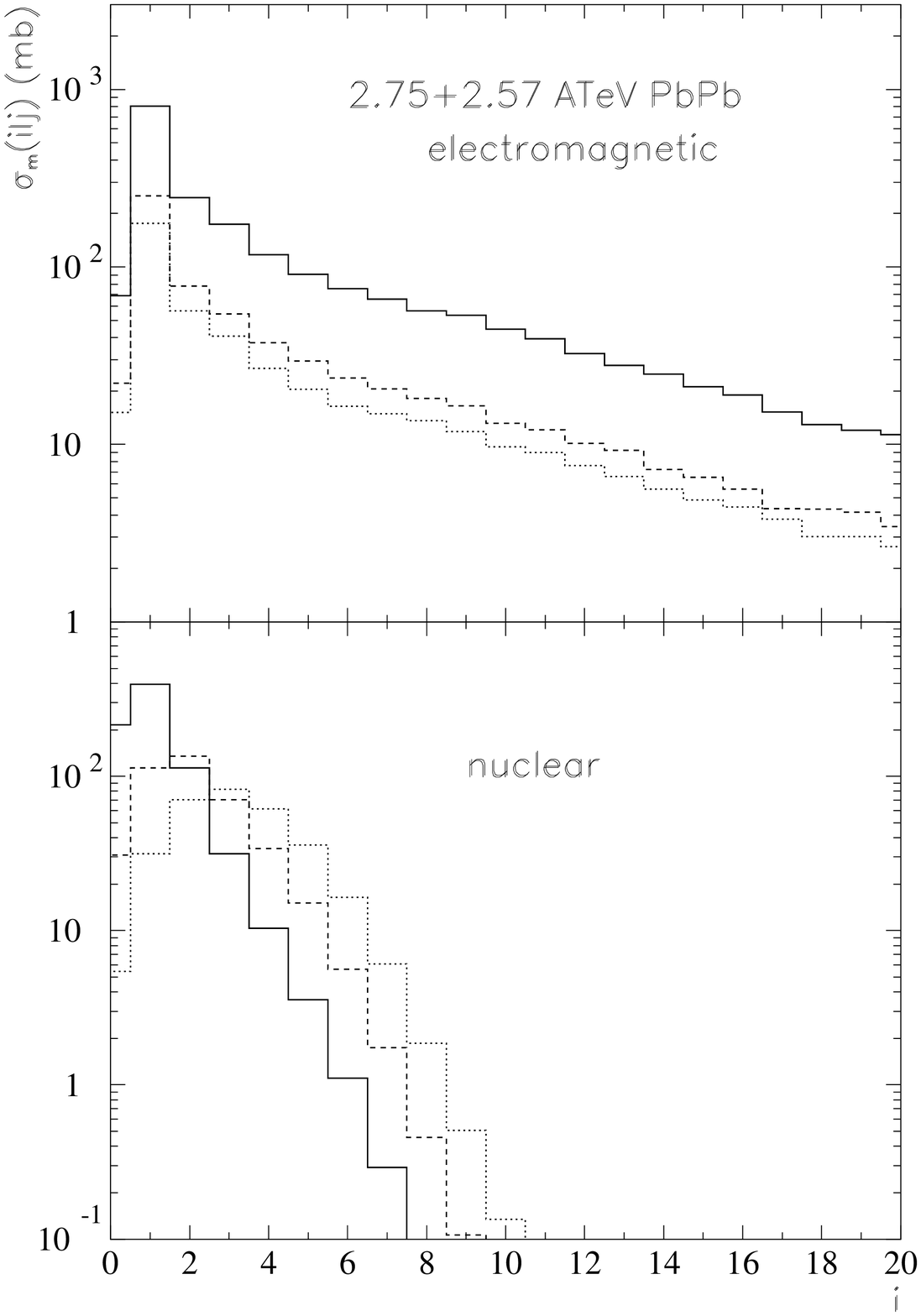}
\caption{The same as in Fig.~\protect\ref{fig:rhic2d}, but for
2.75+2.75 ATeV PbPb collisions at LHC.}
\label{fig:lhc2d}
\end{centering}
\end{figure}

The condition for heavy ion dissociation to be mutual leads to some
specific features for nuclear and electromagnetic interactions.
The former interaction causes mainly symmetric or quasi-symmetric
dissociation. 
The latter makes very probable asymmetric dissociation
like $(1nX\mid 5nY)$ or even $(1nX\mid 10nY)$. 
This feature is shown
in Figs.~\ref{fig:rhic2d} and \ref{fig:lhc2d} where the cross sections 
with one, two or three neutrons in one arm of the ZDC setup are
presented. As one can see, $(1nX\mid 10nY)$ dissociation is almost absent
in nuclear collisions. 
This is not true for electromagnetic dissociation where
the number of $(1nX\mid 10nY)$ events 
is approximately 1--5\% 
of the main dissociation channel, $(1nX\mid 1nY)$.  

In the RELDIS model nuclei undergo dissociation independently
in the electromagnetic fields of each other.
Therefore there is no correlation between the numbers of neutrons,
$n_1$ and $n_2$, emitted by each of the nuclei, and asymmetric
dissociations are possible along with symmetric ones. 
The extreme case of asymmetric dissociation is, of course,
the single dissociation process. 
The nuclear dissociation considered in the framework of the abrasion
model has different characteristic features. 
Namely, the numbers of emitted neutrons and protons are 
correlated due to the condition
$z_1+n_1=z_2+n_2$, see Sec.~\ref{abrmodel}, and nuclear dissociation 
is always mutual. The latter condition was also adopted in 
Ref.~\cite{Baltz2}.

\begin{figure}[htb]
\begin{centering}
\epsfxsize=0.7\textwidth
\epsffile{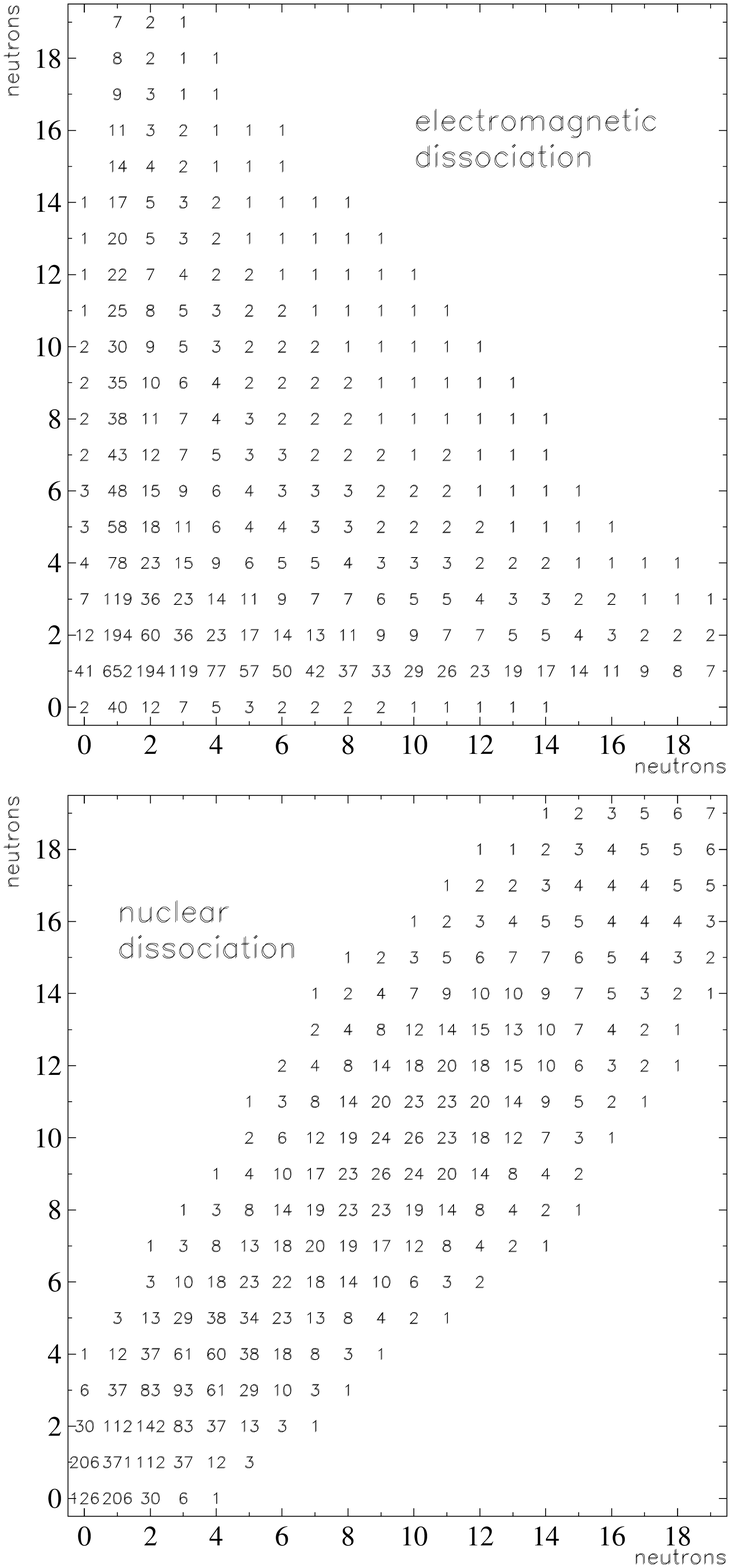}
\caption{Mutual dissociation cross sections (mb) for neutron emission
($i,j=0nX,1nX,...19nX$) due to electromagnetic 
($\sigma^{ED}_m(i\mid j)$, top panel), 
and nuclear ($\sigma^{nuc}_m(i\mid j)$, bottom panel), dissociation 
in 100+100 AGeV AuAu collisions at RHIC.}
\label{fig:rhic}
\end{centering}
\end{figure}

\begin{figure}
\begin{centering}
\epsfxsize=0.7\textwidth
\epsffile{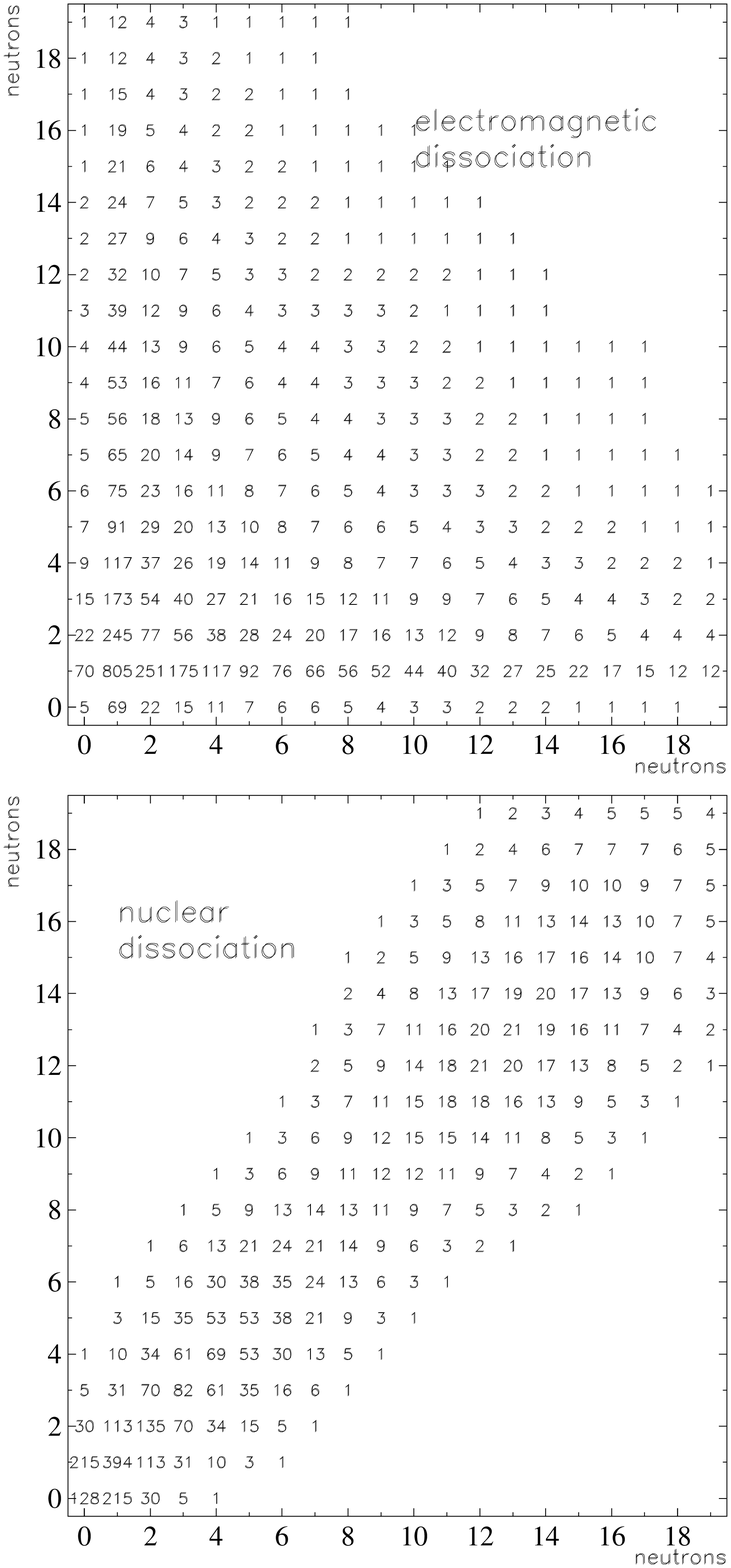}
\caption{Same as in Fig.~\protect\ref{fig:rhic} but 
for 2.75+2.75 ATeV PbPb collisions at LHC.}
\label{fig:lhc}
\end{centering}
\end{figure}

The main results of our study are presented in Figs.~\ref{fig:rhic} 
and \ref{fig:lhc}. They show the electromagnetic and nuclear dissociation 
cross sections. Since $1n$ and $2n$ emission in electromagnetic 
collisions is enhanced due to the GDR and QD absorption mechanisms,
the corresponding strips are prominent in the plots. Simultaneous
GDR excitation in both of the nuclei is a dominant process leading to
the mutual dissociation, but it is responsible only for a part
of the total mutual dissociation cross section, $\sim 18\%$ 
at RHIC, for example.
The rest is provided mainly by asymmetric processes, when one of the
nuclei is excited in a GDR state, while another nucleus absorbs a photon
with the energy above the GDR region,
which leads to emission of many neutrons. 

As seen in Figs.~\ref{fig:rhic} and \ref{fig:lhc}, 
the probabilities of the simultaneous emission of three and more neutrons 
are small and such processes with participation of high energy photons 
are distributed
over the large area in the plots.    

For the sake of completeness the cross sections
of the mutual dissociation without neutrons emitted from one or
both of the collision partners are also included. The rates
of such processes, when mainly the proton 
emission takes place, are small. This is another difference between the
electromagnetic and nuclear dissociation. The nuclear 
interaction events, when only a proton is removed from one or both of 
the nuclei, are very probable, see Figs.~\ref{fig:rhic} and 
\ref{fig:lhc}.

\section{Conclusions}

Since its experimental discovery, the process of electromagnetic
dissociation of heavy ions has been studied only in fixed target 
experiments. In the year 2000 the Relativistic Heavy Ion Collider 
(RHIC) became operational at BNL and among other experiments the 
electromagnetic
dissociation of ultrarelativistic heavy ions can now be 
investigated in collider
kinematics by means of the Zero Degree Calorimeters 
(ZDC)~\cite{Adler}. This makes possible to study the mutual heavy ion 
dissociation process for the first time.

In the present paper the equivalent photon method and the abrasion model
for grazing nuclear collisions were extended
to the case of mutual dissociation of collision partners.
Our numerical results for the total heavy ion dissociation cross
sections are very close to the results of related studies, 
Ref.~\cite{Krauss,Weneser,Baltz2}. Restricting ourselves to the domain of
equivalent photon energies below 24 MeV, where only the 
giant resonance excitations are possible in electromagnetic dissociation,
we obtained the cross section of correlated 1n-1n emission close to one
of Ref.~\cite{Baltz2}. However,
as we have found, at collider energies the neutron emission process 
in mutual electromagnetic dissociation is not entirely exhausted
by the simultaneous excitation and decay of the giant resonances in
both of the colliding nuclei. Apart from the mutual GDR excitation,
asymmetric processes with the GDR excitation in one of the nuclei accompanied
by a photonuclear reaction in the other nucleus are very probable.   
A wide set of photonuclear reactions should be taken into
account to obtain a realistic estimation of the mutual dissociation 
rate. 

As we found, the cross sections of emission of two and especially three
neutrons in mutual dissociation are very sensitive to the
presence of the next-to-leading-order processes where a collision
partner (or even both of the nuclei) absorbs a pair of photons in a
collision event. The contribution of multiple photon absorption 
processes to the total {\em single} electromagnetic dissociation cross section 
at ultrarelativistic energies is 1--2\% only. This is very different from
the case of {\em mutual} electromagnetic dissociation where 30-40\% of
events are due to multiple photon absorption. Indeed, the main part
of mutual dissociation events takes place in close collisions with
small $b\geq b_c$, where the probability to absorb a virtual 
photon is large, and hence, two or more photons can be absorbed
by each of the collision partners.   
         
We have examined the reliability of our results by studying
their sensitivity to the variation of input 
data and parameters. Trying to answer a key question on whether the
mutual dissociation cross section can be calculated with high
accuracy, we have critically reviewed our model assumptions and 
the results of previous theoretical and experimental studies of
photonuclear reactions and heavy ion dissociation processes.

The ambiguity in the calculations of $1n-1n$ correlated  
emission cross section alone, $\sigma (1nX\mid 1nY)$,  
is found to be up to 15\%.  
This is mainly due to the difference in the values of the 
photoneutron cross sections measured in different experiments.  
However, the ambiguity is lower, $\sim 5-7$\%, 
if the sum of one and two neutron 
emission channels, $\sigma (1nX\mid 1nY) +2 \sigma (1nX\mid 2nY)
+\sigma(2nX\mid 2nY)$, is considered. Therefore, 
it is a kind of cumulative neutron emission rate 
which should be used as the luminosity measure at colliders.

We have found several distinctive features of the 
mutual electromagnetic dissociation process which are 
helpful for its experimental identification. 
Beside the enhancement of $1n$ and $2n$ emission channels, 
the electromagnetic interaction 
leads to very asymmetric mutual dissociation channels where
only one neutron is lost by one collision partner while many
neutrons are lost by another partner. Such dissociation pattern
is very unlikely in grazing nuclear collisions with participation of the
strong nuclear forces.  
 
The correlated emission of one or two neutrons 
in both the forward and backward directions without any additional
particles in the mid-rapidity region 
can be used as a clear 
sign of the electromagnetic dissociation of ultrarelativistic 
heavy ions. 
The identification of such mutual dissociation events and counting 
their rates in both arms of ZDC along with the calculation results of 
the present paper can provide a basis for an absolute luminosity 
calibration at RHIC. Similar methods can be used for 
the ALICE heavy ion experiment~\cite{ALICE-ZDC} planned at the 
future Large Hadron Collider (LHC) to be build at 
CERN.

\section{Acknowledgements}

We are grateful to A.J.~Baltz, C.A.~Bertulani, A.S.~Botvina, 
M.B.~Chadwick, G.~Dellacasa, G.~Giacomelli, J.J.~Ga\-ard\-h\o{j}e, 
A.S.~Iljinov, A.B.~Kurepin and M.~Murray for useful discussions. 
Special thanks are due 
to S.N.~White who pointed our attention to the subject of the paper and 
encouraged us in the present study. I.A.P. thanks ENEA and the 
Niels Bohr Institute for the warm hospitality and financial support.
I.N.M. acknowledges the support from the Niels Bohr Institute
and from the grant RFBR-00-15-96590.


\begin{center}
\begin{table}[ht]
\caption{ Total cross sections (barn) 
of single and mutual 
dissociation calculated by the RELDIS code, abrasion model
and by other authors 
for AuAu and PbPb collisions at RHIC and LHC.}
\vspace{0.3cm}
\begin{tabular}{|c|c|l|c|l|}
\hline\hline
                 & Disso-        & \ \ \ \ \ \ First  & Second &  \ \ All    \\
                 & ciation &  \ \ \ \ \ \ order & order  &  \ \ contri- \\
                 & process        &        &        &  \ \ butions \\
\hline\hline
               &Single&         &    &   \\
               &electromagnetic&\ \ \ \ \ \ \ \ 82&\ \ 1.78  &\ \ \ \ 83.8 \\
    & &        &       &   \\
\cline{2-5}
 65+65 AGeV &Mutual&    &   &  \\
 AuAu at RHIC &electromagnetic & \ \ \ \ \ \ \ \ \ -- &\ 2.5 &\ \ \ \ 3.6  \\
               &     &        &       &     \\
\cline{2-5}
 & & & & \\
 &Nuclear& \ \ \ \ \ \ \ \ \ -- & --  & \ \ \ \ 7.29 \\
 & & & &  \\
\hline\hline
                     &Single& \ \ \ \ \ \ 93.2   & 1.86   & \ \ \ 95.1  \\
                     &electromagnetic&  &    &  \ \ \ 88~~\protect\cite{Krauss} \\
    & &        &       &  \ \ \ 95~~\protect\cite{Weneser}\\
\cline{2-5}
 100+100 AGeV &Mutual& $0.39\cdot 10^{-3}$~\protect\cite{BaurHen} &\ 2.6
&\ \ \ 3.8 \\
   AuAu at RHIC &electromagnetic &  $0.49\cdot 10^{-3}$~\protect\cite{Benesh}&  &\ \ \ 3.9~\protect\cite{Baltz2}  \\
               &     &        &       &     \\
\cline{2-5}
 & & & & \ \ \ 7.29 \\
 &Nuclear& \ \ \ \ \ \ \ \ \ -- & --  & \ \ \ 7.09~\protect\cite{Baltz2}\\
 & & & &  \\
\hline\hline
                     &Single  & \ \ \ \ \ \ 212 &\ 3 & \ \ \ 215   \\
                     &electromagnetic&    &       & \ \ \ 214~\protect\cite{Krauss} \\
   &    &        &       &  \ \ \ 220~\protect\cite{Weneser} \\
\cline{2-5}
 2.75+2.75 ATeV & Mutual & $0.43\cdot 10^{-3}$~~\protect\cite{BaurHen}&\ \ 3.9
&\ \ \ 6.2 \\
  PbPb at LHC & electromagnetic & $0.54\cdot 10^{-3}$~~\protect\cite{Benesh}&  & \ \ \ 7.15~\protect\cite{Baltz2} \\
                      &     &        &       &     \\
\cline{2-5}
 & & & & \ \ \  \\
 &Nuclear& \ \ \ \ \ \ \ \ \ -- & --  & \ \ \ 7.88 \\
 & & & &  \\
\hline\hline
\end{tabular}
\label{T1}
\end{table}
\end{center}

\begin{center}
\begin{table}[h]
\caption{Cross sections (barn) for $^{197}{\rm Au}$ dissociation   
induced by 158 AGeV Pb beams.
Theoretical results  are obtained by the RELDIS code
and abrasion model.
Experimental data are taken from
Ref.~\protect\cite{Hill-Petridis}.  RELDIS results without direct $1n$
emission are given in parentheses.}
\vspace{0.3cm}
\begin{tabular}{|c|c|c|c|c|c|c|}
\hline\hline
 &\multicolumn{2}{|c|}{\ } 
 & \multicolumn{2}{|c|}{\ } & 
 \multicolumn{2}{|c|}{\ } \\
 &\multicolumn{2}{|c|}{$\sigma_1^{ED}(i)+\sigma_2^{ED}(i)$ } 
 & \multicolumn{2}{|c|}{$\sigma^{nuc}(i)$ } & 
 \multicolumn{2}{|c|}{All contributions } \\
Dissociation &\multicolumn{2}{|c|}{\ } 
& \multicolumn{2}{|c|}{\ } & 
 \multicolumn{2}{|c|}{\ } \\
\cline{2-7} 
channel   &     &        &       &      &       &  \\
 & Experi-& RELDIS  & Experi-  & Abrasion & Experi-  & Theory \\
 & ment   & code    & ment     &   model  & ment     &       \\
     &     &        &       &      &       &  \\
\hline\hline
     &     &        &       &      &       &  \\
 $i=1n$ & $26.4\pm 4.0$& 26.96 & $0.3\pm 0.1$ & 0.43 & $26.7\pm 4.0$ & 27.39 \\
 $^{197}{\rm Au}\rightarrow ^{196}{\rm Au}+n$ &  & (25.09) &  & &  & (25.52) \\
     &     &        &       &      &       &  \\
\hline
     &     &        &       &      &       &  \\
$i=2n$ & $4.6\pm 0.7$ & 4.57 & $0.13\pm 0.4$ & $0.13$ & $4.7\pm 0.7$ & 4.70 \\
$^{197}{\rm Au}\rightarrow ^{195}{\rm Au}+2n$ &  & (6.39) &  &  &  & (6.52) \\
     &     &        &       &      &       &  \\
\hline\hline
\end{tabular}
\label{THill}
\end{table}
\end{center}

\begin{center}
\begin{table}[h]
\caption{Mutual electromagnetic dissociation cross sections for
AuAu collisions at RHIC. $X$ and $Y$ denote any particle, except neutron.
${\cal D}$ means any dissociation channel. Calculation results are given (a) 
for the leading order contribution only, (b) for the sum of 
leading order and next-to-leading-order contributions.   
}
\vspace{0.3cm}
\begin{tabular}{|c|c|c|c|}
\hline\hline
 &Cross & & \\
 &section & (a) & (b) \\ 
 &(mb)& LO & LO+NLO \\
 & & & \\
\hline\hline
 & & & \\
 & $\sigma_m^{ED}(1nX\mid 1nY)$& 612 & 659 \\
 & & & \\
65+65 AGeV & $\sigma_m^{ED}(1nX\mid {\cal D})$& 1244 & 1502 \\
 & & & \\
AuAu at RHIC & $\sigma_m^{ED}(2nX\mid {\cal D})$& 330 & 446 \\
 & & & \\
 & $\sigma_m^{ED}(3nX\mid {\cal D})$& 148 & 274 \\
 & & & \\
\hline\hline
 & & & \\
 & $\sigma_m^{ED}(1nX\mid 1nY)$& 607 & 652 \\
 & & & \\
100+100 AGeV & $\sigma_m^{ED}(1nX\mid {\cal D})$& 1257 & 1518 \\
 & & & \\
AuAu at RHIC & $\sigma_m^{ED}(2nX\mid {\cal D})$& 341 & 461 \\
 & & & \\
 & $\sigma_m^{ED}(3nX\mid {\cal D})$& 155 & 284 \\
 & & & \\
\hline\hline
\end{tabular}
\label{TLONLO}
\end{table}
\end{center}

\begin{center}
\begin{table}[h]
\caption{Sensitivity of the partial mutual 
electromagnetic dissociation 
cross sections to the variation of probability of the direct neutron 
emission in $1n$ channel, $P^{dir}_n$,
to the input photonuclear cross sections and to the corrections for  
next-to-leading-order processes. 
Results obtained with GNASH and RELDIS codes are given for 
100+100 AGeV AuAu collisions. Recommended values are presented in boldface. 
Prediction of Ref.~\cite{Baltz2} for $\sigma_m^{ED}(1nX\mid 1nY)$ 
is given for comparison.
}
\vspace{0.3cm}
\begin{tabular}{|c|c|c|c|c|c|}
\hline\hline
& &\multicolumn{2}{|c|}{\ }&\multicolumn{2}{|c|}{\ }\\
& $E_\gamma\leq 24$ MeV &\multicolumn{2}{|c|}{$E_\gamma\leq 140$ MeV }&
 \multicolumn{2}{|c|}{Full range of $E_\gamma$ }\\
& LO &\multicolumn{2}{|c|}{LO }&\multicolumn{2}{|c|}{LO+NLO}\\
& &\multicolumn{2}{|c|}{\ }&\multicolumn{2}{|c|}{\ }\\
\cline{2-6}
Cross section & & & & & \\
(mb) & RELDIS & GNASH & RELDIS & RELDIS & RELDIS \\
 &$P^{dir}_n=0$ & & $P^{dir}_n=0$ & $P^{dir}_n=0$ & $ P^{dir}_n=0.31$ \\ 
 & & & & & \\
\hline\hline
 & & & & & \\
$\sigma_m^{ED}(1nX\mid 1nY)$ & 437 & 430 & 467 & 549 & {\bf 652} \\
 & \ \ \ \ \ 445~\cite{Baltz2} & & & & \\
 & & & & & \\
\hline
 & & & & & \\
$\sigma_m^{ED}(1nX\mid 2nY)$+  &205 & 221 & 262 & 439 & {\bf 388} \\
$\sigma_m^{ED}(2nX\mid 1nY)$\  &  & & & & \\
 & & & & & \\
\hline
 & & & & & \\
$\sigma_m^{ED}(2nX\mid 2nY)$ & 21 & 28 & 38 & 87 & {\bf 60} \\
 & & & & & \\
\hline
 & & & & & \\
$\sigma_m^{ED}(LMN)$ & 663 & 679 & 767 & 1075 & {\bf 1100} \\
 & & & & & \\
\hline\hline
\end{tabular}
\label{TPdir}
\end{table}
\end{center}

\begin{center}
\begin{table}[h]
\caption{ Sensitivity of the mutual dissociation cross section in nuclear
interactions to the variation of $r_o$ parameter of the 
nuclear density distribution.
Results of the abrasion model are given for 100+100 AGeV AuAu
collisions.  Recommended values are given in boldface.
}
\vspace{0.3cm}
\begin{tabular}{|c|c|c|c|c|}
\hline\hline
 &\multicolumn{4}{|c|}{\ }  \\
 &\multicolumn{4}{|c|}{Abrasion model}  \\
Cross section &\multicolumn{4}{|c|}{\ }   \\
\cline{2-5}
(mb) & & & & \\
 & $r_o=1.09$ & $r_o=1.12$ & $ r_o=1.14$ & $r_o=1.16$ \\
 & $R_o=6.34$ fm & $R_o=6.52$ fm &$R_o=6.63$ fm &$R_o=6.75$ fm \\
 & & & & \\
\hline\hline
 & & & & \\
$\sigma_m^{nuc}(1nX\mid 1nY)$   & 361 & 364 & {\bf 371 }& 382 \\
 & & & & \\
\hline
 & & & & \\
$\sigma_m^{nuc}(1nX\mid 2nY)$+  & 241 & 232 &{\bf 224 } & 226 \\
$\sigma_m^{nuc}(2nX\mid 1nY)$\  & & & & \\
 & & & & \\
\hline
 & & & & \\
$\sigma_m^{nuc}(2nX\mid 2nY)$   & 148 & 147 & {\bf 142} & 139 \\
 & & & & \\
\hline
 & & & & \\
$\sigma_m^{nuc}(LMN)$           & 750 & 743 & {\bf 737 }& 747 \\
 & & & & \\
\hline\hline
\end{tabular}
\label{Tro}
\end{table}
\end{center}

\begin{center}
\begin{table}[h]
\caption{ Sensitivity of the mutual dissociation cross section in nuclear
interactions 
to the variation of the total nucleon-nucleon cross 
section $\sigma_{NN}$.
The results of the abrasion model are given for 100+100 AGeV AuAu 
collisions.  Recommended values are given in boldface.}
\vspace{0.3cm}
\begin{tabular}{|c|c|c|c|}
\hline\hline
 &\multicolumn{3}{|c|}{\ }  \\
 &\multicolumn{3}{|c|}{Abrasion model}  \\
Cross section &\multicolumn{3}{|c|}{\ }   \\
\cline{2-4}
(mb) & & &  \\
 & $\sigma_{NN}=40$ mb &$\sigma_{NN}=50$ mb 
 & $\sigma_{NN}=60$ mb \\
 & & &  \\
\hline\hline
 & & &  \\
$\sigma_m^{nuc}(1nX\mid 1nY)$ & 370 & {\bf 371} & 374 \\
 & & &  \\
\hline
 & & &  \\
$\sigma_m^{nuc}(1nX\mid 2nY)$+  & 233 & {\bf 224} & 222 \\
$\sigma_m^{nuc}(2nX\mid 1nY)$\    & & & \\
 & & &  \\
\hline
 & & &  \\
$\sigma_m^{nuc}(2nX\mid 2nY)$  & 148 & {\bf 142} & 138 \\
 & & &  \\
\hline
 & & &  \\
$\sigma_m^{nuc}(LMN)$  & 751 & {\bf 737} & 734 \\
 & & &  \\
\hline\hline
\end{tabular}
\label{TNN}
\end{table}
\end{center}

\begin{center}
\begin{table}[h]
\caption{Sensitivity of the mutual dissociation cross section in 
electromagnetic and nuclear
interactions to the critical impact parameter $b_c$.
The results of the RELDIS code and abrasion model are given for
100+100 AGeV AuAu collisions.  Recommended values are given in boldface.
}
\vspace{0.3cm}
\begin{tabular}{|c|c|c|c|}
\hline\hline
 & & &  \\
              & $R_{BCV}=1.27$ & $R_{BCV}=1.34$ & $R_{BCV}=1.41$  \\
Cross section & $b_c=14.45$ fm & $b_c=15.25$ fm & $b_c=16.05$ fm  \\
(mb) & & &  \\
 & & &  \\
\hline\hline
 & & &  \\
$\sigma_m^{ED}(1nX\mid 1nY)$ & 677 &{\bf 652} & 629 \\
 & & & \\
\hline
 & & & \\
$\sigma_m^{ED}(1nX\mid 2nY)$+  & 417 &{\bf 388} & 374 \\
$\sigma_m^{ED}(2nX\mid 1nY)$\    & & & \\
 & & & \\
\hline
 & & & \\
$\sigma_m^{ED}(2nX\mid 2nY)$  & 62 & {\bf 60} & 57\\
 & & & \\
\hline
 & & & \\
$\sigma_m^{ED}(LMN)$  & 1156 &{\bf 1100} & 1060\\
 & & & \\
\hline\hline
 & & &  \\
$\sigma_m^{nuc}(1nX\mid 1nY)$ & 379 & {\bf 371} & 390 \\
 & & & \\
\hline
 & & & \\
$\sigma_m^{nuc}(1nX\mid 2nY)$+  & 240 & {\bf 224} & 259 \\
$\sigma_m^{nuc}(2nX\mid 1nY)$\    & & & \\
 & & & \\
\hline
 & & & \\
$\sigma_m^{nuc}(2nX\mid 2nY)$  & 141 & {\bf 142} & 151 \\
 & & & \\
\hline
 & & & \\
$\sigma_m^{nuc}(LMN)$  & 760 & {\bf 737} & 800 \\
 & & & \\
\hline\hline
 & & & \\
$\sigma_m^{ED}(LMN)$+  & 1916 & {\bf 1837} & 1860 \\
$\sigma_m^{nuc}(LMN)$\  & & & \\
 & & & \\
\hline\hline
\end{tabular}
\label{Tbc}
\end{table}
\end{center}

\clearpage

\newpage

\begin{figure}
\begin{center} 
\vspace{1cm} 
\unitlength=1cm 
\begin{picture}(8.0,2.0) 

\SetWidth{1.0}
\ArrowLine(0,100)(60,100)  
\Text(0,3.9)[]{$A_1$} 
\SetWidth{0.5}  
\GCirc(60,100){4}{1.} 
\SetWidth{1.0}
\ArrowLine(64,100)(120,100)  
\Text(3.1,3.9)[]{$A_1$}   

\SetWidth{0.5}
\Photon(60,97)(60,36){3}{4} 
\Text(1.7,2.2)[]{$E_1$} 

\SetWidth{1.0}
\ArrowLine(0,33)(60,33)  
\Text(0,0.8)[]{$A_2$} 
\SetWidth{0.5}  
\GCirc(60,33){4}{0} 
\SetWidth{2.0}
\ArrowLine(60,33)(120,33)  
\Text(3.1,0.8)[]{$A_2^\star$}
\SetWidth{0.5}  
\GCirc(120,100){4}{0}
\SetWidth{2.0} 
\ArrowLine(120,33)(180,33)  
\Text(5.9,0.8)[]{$A_2^\star$}
\SetWidth{0.5}  
\GCirc(120,33){4}{1.} 
  
\Photon(120,97)(120,36){3}{4} 
\Text(3.9,2.2)[]{$E_2$}

\SetWidth{2.0}
\ArrowLine(120,100)(180,100)  
\Text(5.9,3.9)[]{$A_1^\star$}   
 
\end{picture} 
\end{center} 
\caption{ 
Mutual electromagnetic excitation of relativistic nuclei: second order 
process. Open and closed circles denote elastic and inelastic vertices,
respectively.  
\label{EMsecond}  
} 
\end{figure}
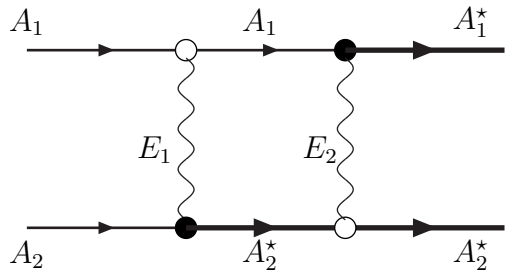 


\end{document}